\documentclass[%
 reprint,
 amsmath,amssymb,
 aps,
]{revtex4-2}

\usepackage{graphicx}
\usepackage{dcolumn}
\usepackage{bm}

\usepackage[utf8]{inputenc}
\usepackage[mathscr]{euscript}
\usepackage{gensymb,amsfonts}
\usepackage{amssymb, }
\usepackage{array}
\usepackage[T1]{fontenc}
\usepackage[margin=1in]{geometry}
\usepackage{lmodern}
\usepackage{tabularx}
\usepackage{fancyhdr}
\usepackage{graphicx}
\graphicspath{{images/}}
\usepackage[utf8]{inputenc}
\usepackage[english]{babel}
\usepackage{amsthm}
\usepackage{braket}
\usepackage{tikz}
\usepackage{subfig, float}
\usepackage[utf8]{inputenc}
\usetikzlibrary{cd}
\usepackage{algorithm}
\usepackage{algpseudocode}
\usepackage{circdraw}
\usepackage{cancel}
\usepackage[thicklines]{easytable}

\newtheorem{proposition}{Proposition}

\newcommand{\half}{\frac{1}{2}}

\newcommand{\supp}{\text{\normalfont Supp}}

\newcommand{\mc}{\mathcal}

\newcommand{\mb}{\mathbb}

\newcommand{\cnot}{\text{CX}}
\newcommand{\h}{\text{H}}
\newcommand{\p}{\text{S}}
\renewcommand{\r}{\text{R}}

\newcommand{\qswap}{\text{SWAP}}
\newcommand{\pfg}{\text{PFG}}

\newcommand{\sgnbt}{\text{signbit}}
\newcommand{\Sp}{\text{Sp}}

\begin{document}

\title{Graph Optimization Perspective for Low-Depth Trotter-Suzuki Decomposition}

\author{Albert T. Schmitz}
 \email{albert.schmitz@intel.com}
\affiliation{\it Intel Labs, Intel Corporation, Hillsboro, Oregon 97124, USA \\
Department of Physics and Center for Theory of Quantum Matter,\\
University of Colorado, Boulder, Colorado 80309, USA
}%

\author{Nicolas P. D. Sawaya}
\affiliation{\it Intel Labs, Intel Corporation, Santa Clara, California 95054, USA
}%

\author{Sonika Johri}
 \altaffiliation{SJ has since moved on from Intel Labs.}
\affiliation{\it Intel Labs, Intel Corporation, Hillsboro, Oregon 97124, USA 
}%

\author{A.Y. Matsuura}
\affiliation{\it Intel Labs, Intel Corporation, Hillsboro, Oregon 97124, USA 
}%

\date{\today}

\begin{abstract}
Hamiltonian simulation represents an important module in a large class of quantum algorithms and simulations such as quantum machine learning, quantum linear algebra methods, and modeling for physics, material science and chemistry. One of the most prominent methods for realizing the time-evolution unitary is via the Trotter-Suzuki decomposition. However, there is a large class of possible decompositions for the infinitesimal time-evolution operator as the order in which the Hamiltonian terms are implemented is arbitrary. We introduce a novel perspective for generating a low-depth Trotter-Suzuki decomposition assuming the standard Clifford+RZ gate set by adapting ideas from quantum error correction. We map a given Trotter-Suzuki decomposition to a constrained path on a graph which we deem the Pauli Frame Graph (PFG).  Each node of the PFG represents the set of possible Hamiltonian terms currently available to be applied, Clifford operations represent a move from one node to another, and so the graph distance represents the gate cost of implementing the decomposition. The problem of finding the optimal decomposition is then equivalent to solving a problem similar to the traveling salesman. Though this is an NP-hard problem, we demonstrate the simplest heuristic, greedy search, and compare the resulting two-qubit gate count and circuit depth to more standard methods for a large class of scientifically relevant Hamiltonians, both fermionic and bosonic, found in chemical, vibrational and condensed matter problems which naturally scale. We find in nearly every case we study, the resulting depth and two-qubit gate counts are less than those provided by standard methods, by as much as an order of magnitude. We also find the method is efficient and amenable to parallelization, making the method scalable for problems of real interest. 
\end{abstract}

\maketitle

\section{Introduction}
In the near-term intermediate scale quantum (NISQ) era of quantum computation, a great deal of work has gone into the compilation and optimization of quantum circuits. Optimization is especially important as near-term quantum hardware is limited by depth and number of gates before noise and other constraints render the outcome unusable. Most current methods are focused on taking as user input a sequence of circuit elements which are then manipulated iteratively via a suite of optimization routines \cite{scaffcc2015, projQ2018, qiskit2019, tket2020, kliuchnikov2013optimization, abdessaied2014quantum, nam2018automated, pointing2021optimizing, xu2022quartz}. Though these methods reduce circuit complexity, they often only do so by finding local patterns and can greatly suffer from non-optimal choices by the user. Thus we look to find a method which exploits a higher-level representation to decrease the depth of the final quantum program and is not prohibitively expensive when applied to real problems, similar to methods found in Refs.\cite{PaykinSchmitz2023PCOAST, Zhang2019, Sivarajah_2021_tket, cowtan2019phase, Lu2021}.

In this paper, we consider a general Hermitian operator of any particle type as one such representation.  An ubiquitous use for a quantum computer approximates the unitary generated by exponentiating such operators. Such a use can be the primary computation or a module within some larger algorithm. This includes problems found in condensed matter \cite{wecker15_correlec, abrams97_fermion}, high-energy physics \cite{Nachman21_hep}, materials science \cite{Bauer20_chemrev, Ma2020_mater}, and chemistry \cite{cao19_chemrev,mcardle20_rev,elfving20_indust,mcardle19_qvibr, sawaya20_vibrspec, ollitrault20_reiher_qvibr}. The most common methods for generating this approximation is the Trotter-Suzuki decomposition \cite{suzuki76}, though extensions and more advanced methods exist \cite{Berry2015, qsp2017,  low2019_qubitization, Childs2019_random, Campbell2019_random, berry15_blackbox, berry15_hamsim_qwalk,childs19_theoryoftrotter}. One first writes the Hermitan operator in a basis consisting of all tensor-products of single-qubit Pauli operators (henceforth referred to as Pauli operators) using several methods depending on the particle type or problem description; see Section \ref{sec:mapping}. The Trotter-Suzuki decomposition then applies multiple repeated {\it Trotter steps}, each of which is a sequence of unitary rotations about each Pauli operator term with an angle whose size is inversely proportional to the accuracy of the approximation. We refer to this type of circuit form as the {\it product-of-Pauli-rotations} form (PoPR).

 Though the PoPR form is natural for qubits and is universal for computation, this expression of a unitary does not map immediately to most hardware implementations of quantum computation. In particular, we focus on implementations for which the native gate set consists of one two-qubit entangling Clifford gate such as the controlled-not (CX) or controlled-Z (CZ) gate and all single-qubit rotation gates, what we refer to as the Clifford+RZ gate set.\footnote{As a practical matter, no implementation can truly realize all arbitrary single-qubit rotations natively. However, one only needs at minimum the Clifford $H$ gate and the non-Clifford $T$ gate, at which point any other rotation can be achieved with arbitrary precision \cite{Nielsenbook}. Thus it is merely a theoretical convenience to assume arbitrary rotations from the outset.} This gate set is known to be universal \cite{Barenco1995}and reflects the gate set of many popular platforms such as semi-conductor quantum dot and superconductor transmon implementations. In both these cases, the two-qubit entangling Clifford gate is noisier and takes longer than any single-qubit rotation, and thus represents the primary limiting resource for these NISQ implementations. The typical translation of the PoPR form into a Clifford+RZ circuit is to implement each Pauli rotation via a so-called CX ladder/staircase \cite{Nielsenbook}, making them expensive individually, and then try to optimize the resulting circuit via Clifford gate optimizations at their intersection. Such optimizations are either local such as those which rely on pattern matching \cite{Fagan2018, Nam2018, Nash2020} or relatively expensive if applied to every sequence of Clifford gates in a circuit\cite{Maslov2018, Aaronson2004, Cowtan2019}. However, all such methods are incapable of choosing an optimal ordering for the application of these rotations, implying that even the best Clifford optimizations can not express many available optimizations. This problem is particularly acute in the case of Trotter-Suzuki decomposition where within a single Trotter step, the order of rotations is arbitrary to within the allowed error of the method. Though there exists ordering methods for specific problem instances\cite{Motta2018, Hastings2015}, one would like a general method. One such methods can be found in Ref.\cite{2022paulihedral} which uses block sorting of the Pauli operators representing the terms based on different considerations.

In this paper, we introduce a perspective and methodology familiar to quantum error correction, but applied to the problem of synthesizing a highly efficient Clifford+RZ circuit from the PoPR form. This perspective allows us to simultaneously choose an efficient ordering of Pauli rotations and efficient Clifford operations which connect these rotations.  We do so by viewing any Clifford+RZ circuit as a walk through a hypothetical graph we deem the {\it Pauli frame graph} (PFG). The PFG contains nodes which correspond to a distinct Pauli frame (similar to the Pauli tableau of Refs. \cite{Maslov2018, Aaronson2004}) where edges are added between nodes if a Clifford gate in our gate set connects the two Pauli frames. Thus we can view any circuit as a walk through this graph, effectively applying each Pauli rotation at some node along the path, where the gate cost of the circuit is now proportional to the length of the walk. Though this graph is never explicitly constructed, it allows us to view optimal circuit synthesis as a graph optimization problem where, despite being an NP-hard problem, we can apply known heuristics. As a demonstration of the power of this view, we implement the simplest  heuristic, the greedy search. In spite of its apparent simplicity, we demonstrate a marked improvement over simple ladder methods for a set of condensed matter and chemistry simulations, both fermionic and bosonic, which have natural scaling properties: The Fermi-Hubbard model, the Bose-Hubbard model, a polyacetylene chain, and a vibronic model. The latter two models have significantly denser Hamiltonians (greater number of terms per qubit) than the former two. We find that in nearly every case our methods result in fewer two-qubit gates and a significantly lower depth. We also demonstrate that the method is computationally efficient in producing the final circuit. 

The remainder of this paper is organized as follows: in Section \ref{sec:ex} we motivate our ideas with a simple example. We follow this with a theoretical description of the PFG, the peripheral details and how a walk on the PFG represents the synthesis of a circuit in Section \ref{sec:PFGthry}. Section \ref{sec:trottercycle} discusses details of this perspective regarding the synthesis of one step in a Trotter-Suzuki decomposition and Section \ref{sec:greedy} outlines the ultra-greedy heuristic algorithm using this perspective. Section \ref{sec:results} describes the models we use to test the ultra-greedy algorithm and the results before making some concluding remarks in Section \ref{sec:conclusion}.

\section{Example and Motivation for the PFG Method}\label{sec:ex}

To motivate the following procedure, we start with a simple example. We assume a basic understanding of the first-order Trotter-Suzuki decomposition. A basic review can be found in Ref. \cite{Nielsenbook}.

Suppose we have a four qubit system and we want to simulate the Hamiltonian,
\begin{align}
H=& \theta_{01} Z_0Z_1 + \theta_{12} Z_1Z_2 + \theta_{23} Z_2Z_3 + \theta_{03} Z_0Z_3 \nonumber \\
  &+\theta_{all} Z_0Z_1Z_2 Z_3.
\end{align}
The time evolution operator generated by this Hamiltonian is,
\begin{align}
U(T)=& \exp\left(-iT H\right) \nonumber \\
       =& \exp\left(-i T \theta_{01}Z_0Z_1\right) \exp\left(-i T \theta_{12} Z_1Z_2\right) \nonumber \\
       & \times \exp\left(-i T \theta_{23} Z_2Z_3\right) \exp\left(-i T \theta_{03} Z_0Z_3\right) \nonumber \\
       & \times \exp\left(-i T \theta_{all} Z_0Z_1Z_2 Z_3\right),
\end{align}
where equality in this case is a consequence of all the constituent operators mutually commuting. Using the standard method of synthesis for this circuit, one generates  $\exp\left(-i T\theta_{01} Z_0Z_1\right)$ for example by computing $Z_0 Z_1$ on qubit 1 via a CX gate, then applying a single-qubit Z rotation, $R_Z$ with angle $2 T \theta_{01}$ and then uncomputing $Z_0 Z_1$. One then moves on to the next term using the same strategy, possibly in parallel, and so on until all terms have been applied. However, we recognize that the uncompute is redundant as  $Z_0 Z_1 Z_2 Z_3= Z_0 Z_1 \times Z_2 Z_3$. As we have computed half this operator after the first step, we should keep $Z_0 Z_1$ until the $\theta_{all}$ term is computed. Thus we are looking to leverage linear dependence in the {\it Pauli space} (to be defined below) and use this dependence to avoid unnecessary uncompute sequences. 

\begin{figure}
\centering
\resizebox{24 em}{8.75 em}{
\begin{tikzpicture}

\qreg{4}{0}{0}{12.5}{.75}

\delim[dashed]{1}{4}{1}{1}

\C{
\X{2}{1.5}
}{1}

\sgate{R_Z(\theta_{01})}{2}{3}

\C{
\X{3}{1.5}
}{4}

\sgate{R_Z(\theta_{23})}{3}{3}

\delim[dashed]{1}{4}{4}{4}

\C{
\X{3}{4.5}
}{2}

\sgate{R_Z(\theta_{all})}{3}{6.2}

\C{
\X{1}{4.7}
}{4}

\sgate{R_Z(\theta_{03})}{1}{6.2}

\delim[dashed]{1}{4}{7.2}{7.2}

\C{
\X{3}{7.7}
}{1}

\sgate{R_Z(\theta_{12})}{3}{9.2}

\delim[dashed]{1}{4}{10.2}{10.2}

\C{
\X{1}{10.7}
}{4}

\C{
\X{2}{11.2}
}{1}

\C{
\X{3}{11.7}
}{2}

\comment{$\begin{pmatrix} Z_0 \\ Z_1 \\ Z_2\\ Z_3 \end{pmatrix}$}{.5}{-.75}

\comment{$\begin{pmatrix} Z_0 \\ Z_0 Z_1 \\ Z_2 Z_3\\ Z_3 \end{pmatrix}$}{3}{-.75}

\comment{$\begin{pmatrix} Z_0 Z_3 \\Z_0 Z_1 \\ Z_0 Z_1 Z_2 Z_3 \\ Z_3 \end{pmatrix}$}{6.2}{-.75}

\comment{$\begin{pmatrix} Z_0 Z_3 \\  Z_0 Z_1 \\ Z_1 Z_2\\ Z_3 \end{pmatrix}$}{9.2}{-.75}

\comment{$\begin{pmatrix} Z_0 \\ Z_1 \\ Z_2\\ Z_3 \end{pmatrix}$}{12.5}{-.75}

\end{tikzpicture}
}

\caption{Circuit diagram for our simple example. Below the circuit represents the frame which provides the effective Z-rotation axis for each qubit.}\label{fig:circ1}
\end{figure}
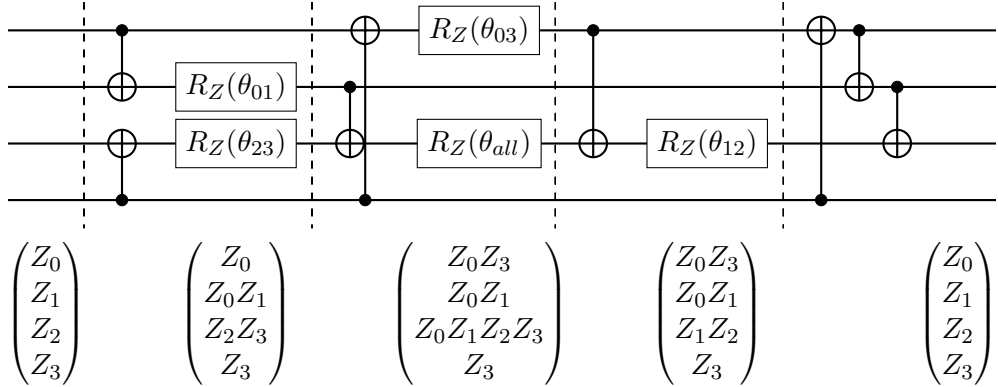

To truly leverage this dependency, we need to be more systematic. The key observation is that each time we apply a CX, we are effectively applying a {\it symplectic automorphism} to the Pauli space. In this case, we only need to consider the space of all Z-operator products which is generated by  or spanned by the basis, also referred to as a {\it frame}, (as defined in Ref. \cite{Bernhard2009}) $(Z_0, Z_1, Z_2, Z_3)$. However, this frame is not unique in generating any Z-type operator. Just as a unitary operator is completely defined by its action on a basis, a symplectic automorphism is uniquely specified by how it transforms the frame. Consider the circuit in Fig. \ref{fig:circ1}. In the first step of the circuit, we are taking $\left(Z_0, Z_1 ,Z_2, Z_3\right) \to \left(Z_0, Z_0 Z_1, Z_2 Z_3, Z_3\right)$. As this is a frame, any Z-type operator can be written as a product (or ``sum'' in the Pauli space) of these operators.  For any given operator, the number of terms in its expansion in this frame (minus one) tells us how may CX gates it takes to compute a rotation for that operator. For example, after the first step in Fig. \ref{fig:circ1}, we have three remaining terms to compute: $Z_1Z_2 =Z_0 \times Z_0 Z_1 \times Z_2$, $Z_0 Z_3=   Z_0 \times Z_2 \times Z_2 Z_3$, and $ Z_0 Z_1 Z_2 Z_3 = Z_0 Z_1 \times Z_2 Z_3$. From this position, it is best to next compute the $\theta_{all}$ term by adding one additional CX gate. From there we can do the same analysis to find that the best move is to include one CX to calculate $Z_0 Z_3$ and one CX to calculate $Z_1 Z_2$. After the final rotation, we are in the frame $\left(Z_0 Z_3, Z_0 Z_1, Z_1Z_2, Z_3\right)$. As we want to return to the original frame, thus completing the Trotter step, we include three additional CXs to return to the $\left(Z_0, Z_1, Z_2, Z_3\right)$ frame. the final circuit uses 8 CX gates and has a depth of $8$ (moving one CX to the left in Fig. \ref{fig:circ1}). Note that if we apply several Trotter steps, we don't need to return to the original frame, but rather start the new Trotter step in the ending Pauli frame of the last step. This leads to a method we refer to as {\it retracing} which we introduce in Section \ref{sec:retrace}.

As one can now see, our circuit is generally traversing different frames, such that every operator of our Hamiltonian is in at least one frame which we discuss below as a path in the PFG. To complete a single Trotter step, we must then come back to the same frame, implying the path is a cycle. This example nicely demonstrates the general idea, but all our terms were mutually commuting which is not general. Instead, we must consider the full Pauli space and not just the Z-type Pauli space. This makes our Clifford automorphisms slightly more complicated, but the general idea is the same.

\section{Description of the PFG and its Uses for Circuit Synthesis}\label{sec:PFGthry}

In this Section we discuss the general theory leading up to a full description of the PFG. This uses many ideas familiar to quantum error correction \cite{Terhal2015} and Clifford circuit simulation. The following draws heavily from and extends results for Refs. \cite{Gottesman1997, Aaronson2004, Maslov2018, Schmitz2019, Tolar2018}. 

\subsection{Pauli Space, Pauli Frames and Symplectic Automorphisms}

Let $N$ be the number of qubits in our system such that the Hilbert space is $\mc H \simeq \mb C_2^{\otimes N}$. For each qubit, we have a natural operator basis in the form of the single-qubit Pauli operators,
\begin{subequations}\label{eq:paulis}
\begin{align}
X =&\begin{pmatrix}
0 &1 \\
1 & 0
\end{pmatrix}, \\
Y =&\begin{pmatrix}
0 &-i \\
i & 0
\end{pmatrix}, \\
Z =&\begin{pmatrix}
1 &0 \\
0 & -1
\end{pmatrix}, 
\end{align}
\end{subequations}
along with the identity. Each Pauli operator is both Hermitian and unitary, implying it squares to the identity and has $\pm1$ eigenvalues. This operator basis can be extended to the full Hilbert space by taking arbitrary tensor products of these single qubit operators. In general, we refer to such tensor products as Pauli operators and denote them only by their single-qubit support, i.e. $Z_0 = Z \otimes I \otimes I \cdots$. Note that all such Pauli operators either commute or anti-commute.

Let $\mc P'$ be the {\it Pauli group}, or set of all tensor products of single-qubit Pauli operators acting on our system which is closed under multiplication. As we often don't care about the overall phases, we further define the {\it Pauli space} as $\mc P= \mc P'/ U(1)$ i.e we have modded out any phase in such a way that, for example, $X_i Z_j \simeq -i X_i Z_j$. Modding out the phase allows us to treat $\mc P$ as a vector space over the field of two elements $\mb F_2$ with dimension $2N$ as given by the fact that the space is generated by all single-qubit Pauli operators and $X_i Z_i \propto Y_i$. To see this forms a vector space, we consider the $N$-qubit identity operator to be the zero element of $\mc P$, addition is given by the product of operators, i.e.  for $p,q \in \mc P$, $pq \to p+q$, and scalar multiplication is the power of the operator, i.e. for $a \in \{0,1\}$, $p^a \to a p$. As $p^2 \simeq I$, $\mb F_2$ is the appropriate field. However, modding out the phase means we have lost the commutation relations between Pauli operators. To recover this, we introduce the non-degenerate symplectic form $\lambda: \mc P \times \mc P \to \mb F_2$ such that 
\begin{align}
(p,q) \mapsto \lambda(p,q) = \begin{cases}
0, & \text{ $p$ and $q$ commute }\\
1, & \text{ otherwise}
\end{cases}.
\end{align}
 One should think of the combination $(\mc P, \lambda)$ much like an inner product space where the inner product form is replaced with a symplectic form ($\lambda(p,p) =0$ for all $p \in \mc P$), thus making this a symplectic vector space. Also much like an inner product space, one generally wants to use $\lambda$ to compute the expansion of any vector in some appropriate basis. This makes such a basis a frame \cite{Bernhard2009}. For a symplectic vector space, the appropriate frame $B_N \subset \mc P$, which we organize as
\begin{align} \label{eq:frame}
B \simeq \begin{pmatrix}
s_0 & \tilde s_0 \\
s_1 & \tilde s_1 \\
s_2 & \tilde s_2 \\
\vdots & \vdots
\end{pmatrix}
\end{align}
which satisfies
\begin{subequations}
\begin{align}
 \lambda(s_i, s_j) =& \lambda(\tilde s_i, \tilde s_j) =0, \\
 \lambda(s_i, \tilde s_j)=& \delta_{ij}.
 \end{align}
 \end{subequations}
Any frame satisfying these relations is referred to as a {\it Pauli frame}. Those familiar with quantum error correction and Clifford circuit simulation will recognize the Pauli frame as equivalent to the Pauli tableau as defined in \cite{Aaronson2004}. In the reference, the left side, un-tilded, operators are refer as {\it stabilizers} and the right side, tilded, operators are referred to as {\it destabilizers}. We label the collection of all such frames as $\mc B$. In particular, one recognizes the single-qubit basis,
\begin{align} \label{eq:origin}
B_0 \simeq \begin{pmatrix}
Z_0 & X_0 \\
Z_1 & X_1 \\
Z_2 & X_2 \\
\vdots & \vdots
\end{pmatrix},
\end{align}
 has this form and is referred to as the {\it origin frame}. Using a given Pauli frame $B$, one can expand any $p \in \mc P$ in this basis according to, 
\begin{align}\label{eq:expand}
p = \sum _i \left( \lambda(s_i, p) \tilde s_i + \lambda(\tilde s_i, p) s_i \right).
\end{align}

Linear maps on the space $(\mc P, \lambda)$ also have a notion similar to unitarity in inner product spaces.\footnote{Recall that unitarity is defined in the context of an inner product space $(\mc V, \braket{,})$ where a unitary operator is defined as any linear operator $U$ such that for all $u,v \in \mc V$, $\braket{U(u), U(v)} = \braket{u,v}$. This then implies the usual notion of unitarity that $U^\dagger = U^{-1}$, where $\dagger$ signifies the adjoint with respects to $\braket{,}$.} We refer to a linear map $\gamma: \mc P \to \mc P$ as a  symplectic automorphism if and only if it preserves $\lambda$. That is, for all $p,q \in \mc P$
\begin{align}
\lambda( \gamma(p), \gamma(q))= \lambda(p,q).
\end{align}
For finite $N$, a symplectic automorphism $\gamma$ can be generated by a member of the {\it unitary Clifford group}, $V_\gamma \in C_N\subset U(2^N)$ which is defined as the normalizer of the Pauli group on the qubit system. The related symplectic automorphism is then given by conjugating all Pauli operators and modding out the phase i.e. $\gamma \simeq  \mc P \to (V_\gamma^\dagger \mc P' V_\gamma)/U(1)$. However, the correspondence is not 1-to-1 as right multiplying a Clifford unitary by any Pauli operator, $q$ and conjugating any other Pauli operator $p$ by it only alters the sign of the result, i.e.
\begin{align}
(V_\gamma q)^\dagger p (V_\gamma q) =  (-1)^{\lambda(p, \gamma^{-1}(q))} V_\gamma^\dagger p V_\gamma, 
\end{align}
 and likewise for left multiplication. Such a sign is then modded out in $\mc P$. However if we keep track of the overall sign of the Pauli operators under this mapping, the resulting map is 1-to-1 with the corresponding Clifford unitary up to an overall phase \cite{Aaronson2004}. This implies the symplectic group on the Pauli space, $\Sp(2N, \mb F_2)$ is equivalent to $C_N/\mc P'$ which we refer to as the {\it Clifford factor group}.  Finally, the  group naturally induces a group action on Pauli frames, $(\Sp(2N, \mb F_2), \mc B_N) \to \mc B_N$ via
\begin{align} \label{eq:frameaction}
\left( \gamma,  \begin{pmatrix}
s_0 & \tilde s_0 \\
s_1 & \tilde s_1 \\
s_2 & \tilde s_2 \\
\vdots & \vdots
\end{pmatrix}\right) \mapsto
\begin{pmatrix}
\gamma(s_0) & \gamma(\tilde s_0) \\
\gamma(s_1) & \gamma(\tilde s_1) \\
\gamma(s_2) & \gamma(\tilde s_2) \\
\vdots & \vdots
\end{pmatrix}
\end{align}
We show in Appendix \ref{ap:group} that this group action is both free and transitive, which importantly implies that for any two frames $B_1, B_2,$ there is a unique symplectic automorphism and thus unique member of the Clifford factor group which connects them. 

To extend Pauli frames so as to encapsulate all Clifford untaries, we only need to include the sign of each Pauli operator contained in the frame. By convention, a member of the Pauli space is mapped to the corresponding member of the Pauli group (i.e. we assign a phase) which corresponds to the conventional Hermitian version in terms of the $X,Y$ and $Z$ operator definitions of Eqs. \eqref{eq:paulis} for each tensor factor of the Hilbert space. Using this convention, we can unambiguously define a sign for all Hermitian Pauli operators. We then view the signed frame $\underline B$ as defining the Clifford unitary $V_{\underline B}$ via the relation,
\begin{align}\label{eq:frametoclif}
 \underline B = V_{\underline B}^\dagger B_0 V_{\underline B},
\end{align}
where conjugation by the unitary  is entry-wise for each member of the signed frame. By our convention, if we map the symplectic automorphism whose group action takes $B_0 \to B$ to the Clifford operator which takes the members of $B_0$ to $B$, where $B$ is nearly $\underline B$ but with all positive signs, then $ V_B^\dagger V_{\underline B}$ is equal to a Pauli operator. This residual Pauli operator, $p$ is such that it flips the signs of $B$ to that of $\underline B$ under conjugation. As $B$ is a basis for the Pauli space, the residual Pauli operators is given uniquely up to a phase by
\begin{align}\label{eq:respauli}
p = \sum_i \left( \sgnbt(\underline s_i) \tilde s_i + \sgnbt(\underline{\tilde s}_i ) s_i \right).
\end{align}

Thus the additional signs resolve the left Pauli operator for the Clifford $V_{\underline B}$ over that of $V_B$ implying we have completely specified the full Clifford operation up to an overall phase.\footnote{A reasonable question one might have is  why we don't include an extra bit in the Pauli space to represent the sign. Though this is possible in principle, this causes two problems. By definition, this would make the binary two-form $\lambda$ degenerate in this extended space, and thus could not be used to form any frame, let alone a Pauli frame. The second reason is that two-qubit Clifford gates such as CX are no longer linear in this extended space. Moreover, this non-linearity is not a practical problem for tracking the sign. See Supplemental Material for details.}

\subsection{Clifford Gate Action on Pauli Frames }

As we assume the Clifford$+\r_Z$ gate set, our Clifford gate set contains  CX, H and $\p= \r_Z\left(\frac{\pi}{2}\right)$, where we note that $\p^2\propto Z$ which is then used to generate all of the $\mc P'$ factor of the Clifford group.\footnote{The choice of CX over CZ is not consequential as we'll see in Section \ref{sec:trottercycle}.} This implies $\p^\dagger =\p^3 \simeq \p$ in $\Sp(2N, \mb F_2)$. Thus we can also use CX, H, and P as the generators for all of $\Sp(2N, \mb F_2)$. We also assume all-to-all connectivity for now and leave methods for incorporating limited connectivity to future work. 

In the case of sequential gate action, each Clifford gate represents two distinct group actions on the frame which we term the {\it backward} and {\it forward} action. That is, if a Clifford gate $g$ is represented by the Clifford operator $V_g$, then the backward and forward group action on the Pauli frame is defined respectively as 
\begin{align}
B \mapsto& V_B^\dagger V_g^\dagger B_0 V_g V_B, \nonumber \\
B \mapsto& V_g B V_g^\dagger,
\end{align}
where we consider the frame-to-Clifford mapping as given in Eq. \eqref{eq:frametoclif}. For the same gate, the backward and forward action result in generally distinct Pauli frames. The forward action applies the conjugation rule (the symplectic  automorphism along with any sign transformation for the signed frame) directly to each member of the frame irrespective of any other.  A backward transformation, on the other hand, first conjugates each member of the origin frame by the adjoint of the gate, and then transforms each of those by the unitary representing the frame. The overall result is a transformation rule which replaces one frame entry with some linear combination of the former entries. From the perspective of symplectic automorphisms, this distinction is artificial as a backward transformation is equivalent to some forward transformation via the relation $V_g^{\text{backward}}= (V_B V_g V_B^\dagger)^{\text{forward}}$. However, the forward transformation is agnostic to the current frame, where as the backward transformation is dependent on the current frame. As we show in a moment, we use the backward transformations for circuit synthesis.\footnote{There is some intuition for why this is. Consider the analogy of rotating 3D vectors. Rotation as an operation on some 3D vector can be view as either the direct or ``forward'' rotation of the vector while fixing coordinates, or the inverse or ``backwards'' rotation of the coordinates while fixing the vector. As we shall see in a moment, the methods we describe below are analogous to the former. We think of the Pauli operators as being fixed, while Clifford operations transform the ``coordinates'' of the space around it.} Alternatively, quantum error correction is typically concerned with the forward action, as the left elements of the Pauli frame/tableau then represent the stabilizers for the state generated by applying the Clifford circuit to the all-zero computational state.

 The backward action of the symplectic automorphism for gate of our gate set is given by the following rules: For any $B=\{(s_i, \tilde s_i)\}_{0\leq i<N} \in \mc B_N$,
\begin{subequations}
\begin{align}
\cnot_{ij}:& s_j\mapsto s_j+ s_i; \nonumber \\
 &\tilde s_i \mapsto \tilde s_i + \tilde s_j, \\
\h_i: &s_i \leftrightarrow \tilde s_i, \\
\p_i : & \tilde s_i \mapsto s_i +\tilde s_i,
\end{align}
\end{subequations}
where members of $B$ not mapped by these rules are unchanged. We shall also identify another important symplectic automorphism, the SWAP, whereby
\begin{align}\label{eq:swap}
\qswap_{ij} :& s_i \leftrightarrow s_j;\nonumber \\
& \tilde s_i \leftrightarrow \tilde s_j \nonumber\\
&= \cnot_{ij}\circ\cnot_{ji}\circ \cnot_{ij} \nonumber \\
&= \cnot_{ji}\circ\cnot_{ij}\circ\cnot_{ji}.
\end{align}
The SWAP generates the symmetric subgroup $\mc S_N \subset \Sp(2N, \mb F_2)$ which corresponds to qubit swapping.

Rules for determining the resulting sign for these gates can be found in Appendix~\ref{ap:signs}.

\subsection{Circuit Synthesis using the PFG}

\begin{figure}[t]

\centering

\includegraphics[scale=.45]{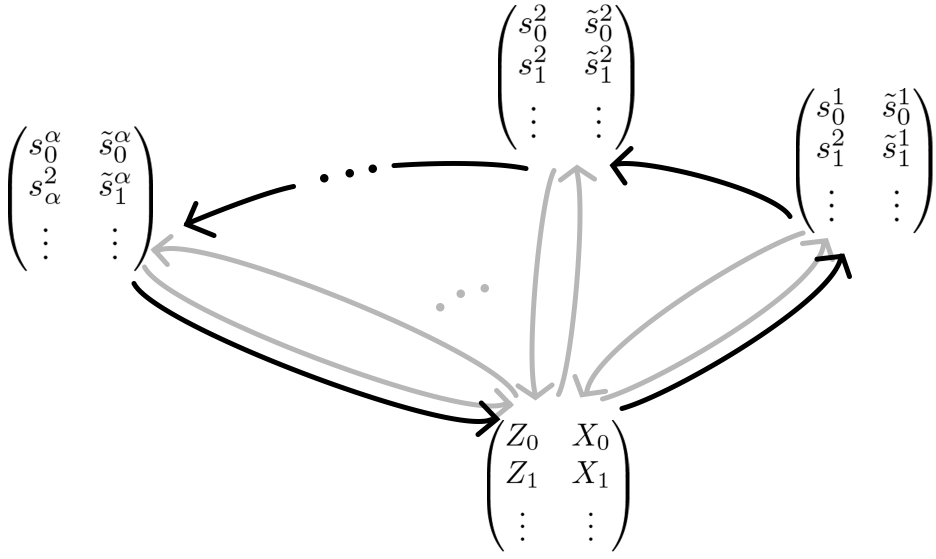}

\caption{Simple depiction of a closed path in the PFG which might be used to realize a unitary in the PoPR form. Grey arrows represent the path generated by standard methods of compute/uncompute, whereas the black arrows represent a more direct path connecting different Pauli frames. }\label{fig:cycle}

\end{figure}

Using the Clifford$+\r_Z$ gate set to generate the Clifford factor group, we can now define the {\it Pauli frame graph }, $\pfg_N=\left(\mc B_N, \mc E_N\right)$ on $N$ qubits as the graph such that the Pauli frames of $\mc B_N$ represent the vertices and the edges $\mc E_N \subseteq \mc B_N \times \mc B_N$ are such that $(B_1, B_2) \in \mc E_N$ if and only if $B_2 = \cnot_{ij}(B_1), B_2 = \h_i( B_1)$ or $B_2 = \p_i(B_1)$ for some indices $i,j$, where we are implicitly using the backward gate action. As every generator is its own inverse, this is an undirected graph. Any directed path on this graph represents an overall symplectic automorphism whose action transforms its starting point to its end point. Because the symplectic group acts freely and transitively on $\mc B_N$,  any two paths between two frames represent an equivalent action on $\mc P$. As we can always map a path to some Clifford circuit, {\bf two paths connected at their endpoints can be mapped to two Clifford circuits which are equivalent modulo a Pauli operator.}  Furthermore, the graph distance between two points is the total unweighted gate cost of the Clifford circuit. Weights can then be added to the edges of $\pfg_N$ to account for unequal gate costs as discussed below.

For our purposes, we are not looking to perform a Clifford unitary as this is not universal and can be efficiently simulated on classical computers \cite{Aaronson2004}. Rather we look to use the Clifford unitaries to achieve a sequence of rotations of the form,
\begin{align}\label{eq:uni1}
U= \prod_\alpha \exp\left(-i \frac{\theta_\alpha}{2} p_\alpha \right),
\end{align}
where $p_\alpha \in\mc P'$ is Hermitian and $\theta_\alpha \in \left(-\pi, \pi\right]$. The presentation of a unitary in this way is referred to as the {\it product-of-Pauli-rotations} (PoPR) form. The PoPR form is universal for computation. To perform any rotation of the form $\exp\left(-i \frac{\theta_\alpha}{2} p_\alpha\right)$, one must be in a Pauli frame such that $p_\alpha \in B_\alpha$, whereby a single qubit operator can be used to apply the rotation. In particular,
\begin{align}
\exp\left(-i\frac{\theta_\alpha}{2} p_\alpha\right ) =& V_{\alpha}^\dagger \exp\left(-i \frac{\theta_\alpha}{2} Z_{j_\alpha}\right) V_{\alpha} \nonumber\\
 &\iff  Z_{j_\alpha} = V_\alpha p_\alpha V_\alpha^\dagger.
\end{align}
 The Clifford $V_{\alpha}$  represents a Clifford automorphism $\gamma_{\alpha}(B_0) = B_\alpha,$ and a path in $\pfg_N$ such that $B_0 \to B_\alpha$ and $V_{\alpha}^\dagger$ represents the opposite path from $B_\alpha \to B_0$. Between the application of two adjacent angles $\theta_{\alpha}$ and $\theta_{\alpha +1}$ we have the product $V_{ \alpha+1} V_{\alpha}^\dagger$. This operator is also represented by a path in $\pfg_N$ such that $B_\alpha \to B_0 \to B_{\alpha+1}$. But as any path performs almost the same Clifford unitary, we can dramatically reduce the gate cost (graph distance) by finding a direct path from $B_{\alpha} \to B_{\alpha+1}$, then mapping it to a Clifford unitary $V'_{\alpha \to \alpha + 1}$ which must be equivalent to $V_{\alpha+1} V_{\alpha}^\dagger$ up to a Pauli operator $p_\text{res}$ as depicted in Fig. \ref{fig:cycle}. Furthermore, since $p_{\alpha +1}$ and $p_\text{res}$ either commute or anti-commute, we can always push $p_\text{res}$ past our rotation at the cost of a sign for our rotation angle, i.e.
\begin{align}
\exp(-i \theta_{\alpha+1} p_{\alpha+1}) p_\text{res}= p_\text{res}\exp(-i
 \sigma \theta_{\alpha+1} p_{\alpha+1}),
\end{align}
 where $\sigma =(-1)^{\lambda(p_{\alpha+1}, p_\text{res})}$. $p_\text{res}$ can then be mapped to some other Pauli on the other side of $V'_{\alpha+1 \to \alpha+2}$ and so on, in which case we can push all these Pauli operators to the end of our circuit. Thus our total unitary is given by
\begin{align}\label{eq:circredux}  
U= p_\text{res} V'_{M \to 0} \left(\prod_{\alpha=1}^M  V'_{\alpha \to \alpha + 1} \exp\left(\pm i \theta_\alpha Z_{j_\alpha}\right)\right) V'_{0 \to 1} 
\end{align}
where $p_\text{res}= V'_{M \to 0} \left(\prod_{\alpha=1}^M  V'_{\alpha \to \alpha + 1}\right) V'_{0 \to 1}$ is the {\it residual Pauli operator} and can be extracted if we considered the signed Pauli frame.\cite{ PaykinSchmitz2023PCOAST}. We can view the entire process as a path through $\pfg_N$ which contains all $B_\alpha$ in order and, as it ends at $B_0$, forms a closed cycle. 

The real power of this viewpoint is not just as a means of replacing $V_{ \alpha+1} V_{\alpha}^\dagger$ with some $V'_{\alpha \to \alpha + 1}$ as a reasonably good Clifford circuit optimization protocol can do this. Instead, we use the above to justify a flipped view on circuit synthesis. Instead of assuming the order as we have in Eq. \eqref{eq:uni1}, we allow the order to be permuted based on the use-case. Then circuit synthesis performs a walk through $\pfg_N$ where at every step, we have all members of the Pauli frame available to be applied, based on Eq. \eqref{eq:frametoclif}. Thus the real power of this perspective is the ability to choose a more optimal ordering of the rotations, based upon the availability of rotations to be effectively applied in a given frame while simultaneously finding efficient Clifford circuits to connect them.

Though this process can be used for general purpose circuit optimization as discuss in Ref.~\cite{2023Paykin}, we focus on the use case where we wish to approximate a unitary $U= \exp(-iTH)$ with Hamiltonian $H$ for a time $T$ using the Trotter-Suzuki decomposition. We can always expand our infinitesimal Hamiltonian in the Pauli operator basis(see Section \ref{sec:mapping}) as 
\begin{align}
\delta H= \sum_{\alpha=0}^{M} \theta_\alpha p_\alpha,
\end{align}
 where $\theta_\alpha \ll 1$. So long as the angles are small enough, error due to the order of rotations in Eq. \eqref{eq:uni1} is within our error tolerance and so we are allowed to choose the order to minimize our cycle in $\pfg_N$. For Trotter-Suzuki decomposition, the problem of $\pfg_N$ circuit synthesis is stated as follows: {\bf Find a cycle $\{B_\beta\}_{0\leq \beta< N}$ in $\pfg_N$ such that for every $\alpha$ there exists a $\beta$ such that $p_\alpha \in B_\beta$ in any order.} We refer to any cycle $\{B_\beta\}_{0\leq \beta< N}$ which satisfies this condition as a {\it Trotter cycle}. We then wish to minimize the length of the Trotter cycle in $\pfg_N$. 
 
\section{General Discussion of the Shortest Trotter Cycle Problem}\label{sec:trottercycle}

When looking to solve the shortest Trotter cycle problem, we recognize the similarity to the Traveling Salesman Problem (TSP), if we can efficiently compute the distance between Pauli frames. However, this comparison is not exact because the set of frames we have to visit in the cycle is not unique. To apply a rotation for $p \in \mc P$, any frame containing $p$ is sufficient. So using a similar analogy as that used to describe the TSP, we imagine a new problem which we call the {\it Traveling Shopper Problem} (TShP): Suppose we have a shopper with a list of items to buy. In general, no store sells all the items and many stores sell the same item. The shopper must travel to several stores, buy every item on the list and return home. Furthermore, at any given place, the shopper can cheaply calculate the shortest distance to {\it a} store which sells a given item. The problem is then to find the sequence of stores with the shortest travel distance such that we buy all items and return home. 

The class of problems described by TShP clearly contains TSP as we get the latter from the former by adding the promise that each item on the list is sold by a unique store. This implies that solving TShP exactly is at least NP-hard. Still, we can adapt similar heuristics as those used for TSP.

However before we can discuss a heuristic solution to this problem, we need to be able to calculate the distance as given in the problem statement. That is, from any frame, we need to be able to efficiently calculate the distance from the current frame to any frame which contains a given Pauli operator. To do this, we are going to effectively add weights to the edges of $\pfg_N$. We assume that for NISQ devices, single-qubit Clifford operations are essentially free and so edges associated to $H$ and $P$ gates have zero weight. Furthermore,  we shall treat SWAP gates as effectively having zero weight since we don't care which qubit we apply the $\r_Z$ gate for a given rotation. Thus we define the non-entangling automorphisms as $\cancel E_N= \braket{\qswap_{ij}, \h_i, \p_i}_{0\leq i,j<N}$ and a set of equally entangled frames (EEF) about $B$ as those connected by a member of $\cancel E_N$, i.e. 
\begin{align}
[B]= \{B' \in \mc B_N | B' = \gamma(B), \text {for some } \gamma \in \cancel E_N\}.
\end{align}
As $\cancel E_N$ is a subgroup of $\Sp(2N,\mb F_2)$, the set of all EEFs represents a partitioning  of $\pfg_N$ into subgraphs, and each of these subgraphs are connected by some number of CXs. We can then define the coarse-grained graph, $[\pfg_N]$, as the quotient graphs with respects to the EEF partitioning, i.e. two coarse-grained vertices $[B_1]$ and $[B_2]$ are connected in $[\pfg_N]$ if and only if there exists members  $B'_1 \in [B_1]$ and $B'_2\in [B_2]$ connected by a CX.  

So by finding a Trotter cycle in $[\pfg_N]$ instead of $\pfg_N$, we have effectively imposed our weight function on gates in the gate set. This also justifies our use of the {\it relative support}, $\supp: \mc P\times \mc B \to \mb N$,
\begin{align}
(d,B) \mapsto \supp( p, B)= \sum_i \lambda(s_i,p) \vee \lambda(\tilde s_i, p),
\end{align}
where $B=\{(s_i, \tilde s_i)\}_{0\leq i < N} $ and $\vee$ is the logical OR. This is a generalization to the qubit support of a given Pauli operator, where here we are giving the support of the Pauli $p$ relative to the frame $B$. We argue that relative support is exactly the ``distance''\footnote{It should be noted that $\supp$ is not a distance function in the usual sense for one major reason: the two entries of its domain are not in the same space. Still, Proposition \ref{prop1} provides all the analogous properties to an actual distance function.} in $[\pfg_N]$ we are interested in as described in TShP. In particular, it has the following properties: 

\begin{proposition}\label{prop1}
The relative support satisfies the following properties for all $p,q \in \mc P$ and $B\in \mc B$:
\begin{enumerate}
\item $\supp(p,B) \geq 0$ and $\supp(p,B)=0$ if and only if $p=I$,
\item $\supp(p+q ,B) \leq \supp(p,B) + \supp(q,B)$,
\item $\supp(p, \gamma(B)) = \supp(p,B)$ for all $\gamma \in \cancel E_N$,
\item$\supp(p, B)-1$ is the minimum distance in $[\pfg_n]$ to reach another frame $B'$ which contains $p$.
\end{enumerate}
\end{proposition}

We prove these properties in Appendix~\ref{ap:prop1}. In general, calculating $\supp(p,B)$ scales as $\mc O(N^2)$ as calculate $\lambda$ is $\mc O(N)$. However, as we discuss below, we don't need to calculate the relative support directly for each step of the search, but rather we can update the binary expansion of a Pauli operator in the current frame, in which case calculating the relative support is $\mc O(N)$. We shall use this in the ultra-greedy heuristic discussed below.

\subsubsection{Two-qubit Entangling Gates}

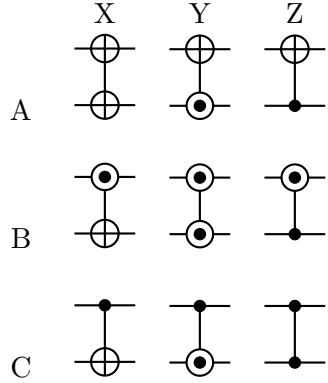
\begin{figure}
\centering
\begin{tabular}{rccc}
& X & Y &Z \\
  A
 \bigskip &
\begin{tikzpicture}
\qreg{2}{0}{0}{.3}{.75}
\A{
\X{2}{.4}
}{1}
\end{tikzpicture}
&
\begin{tikzpicture}
\qreg{2}{0}{0}{.3}{.75}
\A{
\Y{2}{.4}
}{1}
\end{tikzpicture}
&
\begin{tikzpicture}
\qreg{2}{0}{0}{.3}{.75}
\A{
\Z{2}{.4}
}{1}
\end{tikzpicture} \\

 B
\bigskip
&
\begin{tikzpicture}
\qreg{2}{0}{0}{.3}{.75}
\B{
\X{2}{.4}
}{1}
\end{tikzpicture}
&
\begin{tikzpicture}
\qreg{2}{0}{0}{.3}{.75}
\B{
\Y{2}{.4}
}{1}
\end{tikzpicture}
&
\begin{tikzpicture}
\qreg{2}{0}{0}{.3}{.75}
\B{
\Z{2}{.4}
}{1}
\end{tikzpicture}\\

 C
\bigskip &
\begin{tikzpicture}
\qreg{2}{0}{0}{.3}{.75}
\C{
\X{2}{.4}
}{1}
\end{tikzpicture}
&
\begin{tikzpicture}
\qreg{2}{0}{0}{.3}{.75}
\C{
\Y{2}{.4}
}{1}
\end{tikzpicture}
&
\begin{tikzpicture}
\qreg{2}{0}{0}{.3}{.75}
\C{
\Z{2}{.4}
}{1}
\end{tikzpicture}

\end{tabular}

\caption{Table of the nine 2qE gates.}\label{fig:2qEdef}
\end{figure}

Even though $\cancel E_N$ is a subgroup of $\Sp(2N, \mb F_2)$ and ultimately represents a subgroup of the Clifford group, it is not a normal subgroup and as a consequence, $\cnot_{ij}$ acting on members $B, B' \in [B]$ of the same EEF may be taken to distinct EEFs.  This may seem counter-intuitive, but they may take you to different EEFs as a consequence of our definitions. To enumerate the number of distinct two-qubit entangling (TQE) gates, we start by considering only two qubits. Any TQE gate can then be decomposed as some member of $\cancel E_2$ followed by a CX gate and then some other member of $\cancel E_2$. However, any members of $\cancel E_2$ after the CX does not change the EEF and we can mod such operations out in order to count distinct gates which connect different EEFs. Because of this, we can always push a SWAP to the right of the CX and thus we only need to consider pairs of single-qubit operators acting before the CX. Each qubit has six distinct single qubit Clifford gates(modding by single qubit Pauli operators), $I$, $H$, $P$, $HP$, $PH$ and $K\equiv HPH$. Note $K$ is the phase gate for the $\pm$ basis. However, as $P$ commutes with the control and $K$ commutes with the target, we can always pushes such operators to the right and mod them out, thus cutting the number of relevant single qubit operators in half for each qubit. This leaves $3^2=9$ distinct TQE gates for each qubit pair which can be organized into an array as shown in Fig.~\ref{fig:2qEdef}. Generalizing this to multiple qubits, this implies the coordination number for each vertex of $[\text{PFG}_N]$ is $\frac{9}{2} N(N-1)$, i.e nine TQE gates for every qubit pair. 
 Using CX as the example, we use 
\begin{tikzpicture}[thin, scale=0.6, every node/.style={scale=0.6}]
\qreg{1}{0}{0}{.2}{.5}
\X{1}{.3}
\end{tikzpicture}
to represent the $X$-type Pauli operator and 
\begin{tikzpicture}[thin, scale=0.6, every node/.style={scale=0.6}]
\qreg{1}{0}{0}{.2}{.5}
\Z{1}{.35}
\end{tikzpicture}$=$
\begin{tikzpicture}[thin, scale=0.6, every node/.style={scale=0.6}]
\qreg{1}{0}{0}{1.5}{.5}
\sgate{H}{1}{.5}
\X{1}{1}
\sgate{H}{1}{1.5}
\end{tikzpicture}
to represent the $Z$-type Pauli operator. Similarly we introduce the notation of  
 \begin{tikzpicture}[thin, scale=0.6, every node/.style={scale=0.6}]
\qreg{1}{0}{0}{.2}{.5}
\Y{1}{.35}
\end{tikzpicture}$=$
\begin{tikzpicture}[thin, scale=0.6, every node/.style={scale=0.6}]
\qreg{1}{0}{0}{1.5}{.5}
\sgate{P^\dagger}{1}{.5}
\X{1}{1}
\sgate{P}{1}{1.5}
\end{tikzpicture}
to represent the Y-type Pauli operator. These definitions then extend to the TQE gates. As for the names of the TQE gates, we use the convention of labeling the rows by $A, B, C$ and the columns by $X, Y, Z$ as shown in Fig.~\ref{fig:2qEdef}. So for example  $\cnot_{01}$, $\text{CY}_{01}$ and $\text{CZ}_{01}$ are the usual control-X, control-Y and control-Z operations while $\text{AZ}_{0,1}$ and $\text{BZ}_{01}$ are the first two gates in the opposite direction. By convention, we also always order the qubit indices from lowest to highest. We shall refer to these row and column labels as the type similar to that of the Pauli type i.e. CX is of $Z$-type on the first qubit and $X$-type on the second qubit; BY is of $Y$-type on both qubits.

To demonstrate that the 9 operators in Fig. \ref{fig:2qEdef} are distinct edges in $[\text{PFG}_n]$, consider acting on the two qubit origin frame with CX:
\begin{align}
\begin{pmatrix}
Z_0 & X_0\\
Z_1 &  X_1\\
\end{pmatrix}  \xrightarrow{CX}
\begin{pmatrix}
Z_0 & X_0X_1\\
Z_0 Z_1 &  X_1\\
\end{pmatrix} .
\end{align}
We then consider acting on the origin frame with $CZ$ instead
\begin{align}
\begin{pmatrix}
Z_0 & X_0\\
Z_1 &  X_1\\
\end{pmatrix}  \xrightarrow{CZ}
\begin{pmatrix}
Z_0 & X_0Z_1\\
Z_1 &  Z_0  X_1\\
\end{pmatrix} .
\end{align}
Now if the two resulting frames were in the same EEF, then any member of $CZ(B_0)$, say $Z_1$ should be such that $\supp(Z_1, CX(B_0)) =1$, but in fact $\supp(Z_1, CX(B_0))=2$, so these two frames can not be in the same EEF. A similar method can be used to show the other seven TQE operators are also distinct from each other as well as CX and CZ.

 \section{Overview of the Ultra-Greedy Heuristic for Finding a high-efficiency Trotter Cycle}\label{sec:greedy}

We now look to use our knowledge of $[\text{PFG}_n]$ to develop a simple heuristic for finding a Trotter cycle. The simplest heuristic one can apply to such a problem is that of a greedy strategy where at each step, we target one of the nearest Pauli operators in our list of rotations to apply. However, due to the possibility of multiple Pauli operators with the same minimum relative support and non-uniqueness of a frame which can achieve any such rotations, we need to consider how we make decisions in our greedy approach to moving through $[\pfg_N]$. In keeping with the simplest possible heuristic, we also use a greedy method to choose between these possible paths, thus we refer to this strategy as the {\it ultra-greedy} heuristic which is outlined as follows: 

For a given step, of the remaining terms, we find those Pauli operators with the minimum relative support greater than 1. We can calculate this cheaply as it is given by the support of our stored binary vectors. For each of these minimum-distant Paulis, we take every pair of qubits on which it is supported relative to the frame. The support on two qubits consists of four classical bits and there must be non-zero support on each of the two qubits. This leaves only nine possible configurations of the four bits, and for each of these, we argue in Appendix~\ref{ap:signs} that only four of the nine TQE gates can reduce the support by one. This gives us a list of possible gates to apply which will always reduce the support of at least one of these minimum-distance Pauli operators by one. Of these, we need some method of deciding which to apply. We consider two possibly competing considerations. The first is how the gate affects the support of the other remaining terms. So of the remaining terms (including the minimum distant terms) we can calculate the average change in support for the remaining terms after the gate is applied and use this as part of a cost metric. The second consideration is choosing a gate which minimizes the circuit depth. This requires that we have a rough schedule for the gates. Assume all gates take the same amount of time, we keep track of the time of the last gate action for each qubit using ASAP scheduling. For a given gate under consideration, we schedule that gate and find the difference between the scheduled time and the latest scheduled gate (leading temporal edge of the circuit so far). We refer to this difference as the ``pace'' of the scheduled gate. So the cost metric for our possible gate $g$ is
\begin{align}\label{eq:cost}
\text{cost}_{c}(g) =\braket{\Delta \supp_g}  - c|\text{pace}(g)|,
\end{align}
where $\braket{\Delta \supp_g}$is the average change in support of the remaining Paulis after applying $g$ and $c>0$ is a free parameter of the method we refer to as the {\it parallelization credit}. We then choose the gate which minimizes this cost metric. Once a gate is chosen, it is added to the output circuit and all binary vectors are updated. Additionally, the rotations are added to the output circuit at the top of this procedure whenever a Pauli operator with unit support is found. In such a case, we then add an $R_Z$ to the circuit for a local qubit support of $(0,1)$, $R_X$ for $(1,0)$ and $R_Y$ for $(1,1)$, at which time the term is removed from the list of terms to be applied. The algorithm proceeds until all rotations have been applied. The search necessarily halts as every TQE gate we choose reduces the support of at least one term by one, thus always bring us closer to terminating. A more detailed outline of the ultra-greedy search heuristic is given in Appendix~\ref{ap:algo}. 

\subsection{Completing a Trotter Cycle via Retracing}\label{sec:retrace}

The ultra-greedy search only considers synthesizing a single Trotter step which must complete the Trotter cycle by returning to the origin frame. However, when applying several Trotter steps, this is not necessary. Once our greedy search method finds a complete {\it Trotter path} i.e. a path through PFG such that every term appears in at least one frame but does not return to the origin, we are now free to start the next Trotter step in the ending frame. One could restart the search, but it seems reasonable that an efficient path for the next Trotter step is to simply apply the same circuit of the previous step in reverse order, a method we refer to as {\it retracing} as we are simply retracing our steps in the PFG.\footnote{Note that retracing does not undo the work done by the former step because we are not reversing the signs of the rotation angles.} The advantage of retrace is two-fold: First, we avoid the costly origin return path, which can often be a serious downside to greedy search. Second, we automatically obtain the second-order Trotter-Suzuki decomposition, increasing the accuracy of the circuit in approximating the desired unitary. If retracing is undesired, one can also synthesize a return path using Clifford synthesis, \cite{Maslov2018, Aaronson2004} where Ref.~\cite{PaykinSchmitz2023PCOAST} uses methods very similar to ones used here for this purpose.

\subsection{Time Scaling for PFG Ultra-greedy Search}

To assess the performance of the ultra-greedy search heuristic, we consider the expected time scaling with respects to number of qubits, $N$ and number of Pauli terms in the Hamiltonian, $\#(H)$. In the worst case, the algorithm takes at most $N$  iterations through the main loop to apply at least one rotation. However, in practice we expect the number of iterations to be closer to a constant,  in which case the total number of iterations of this main loop should scale as the number of terms. Within that loop, we have two parts: finding the minimum distance terms while simultaneously looking for rotations to apply, and finding the minimum cost TQE gate. In general, finding the minimum cost gate is more costly. Worst case, we need to consider every possible TQE gate of which there are $\mc O(N^2)$. We then need to evaluate the change in support for every remaining term which would require $\mc O(N \#(H))$, but since each gate can only change the support locally, we can use this to reduce this time cost to $\mc O(\#(H))$. Thus we expect the time scaling to be closer to $\mc O( N^2 \#(H)^2)$ with a worst case scaling of $\mc O( N^3 \#(H)^2)$.  This process is also amenable to parallelization to reduce this time. Details of parallelizing the algorithm can be found in Appendix~\ref{ap:parallel}.

\section{Results for the PFG Ultra-greedy Search Algorithm for finding a Trotter Path}\label{sec:results}

In this Section, we demonstrate the use of the PFG ultra-greedy algorithm on four physical models of real-world interest. These models were chosen for their natural scaling behavior. 

\subsection{Physical Examples and their Hamiltonian Mapping to Qubits}\label{sec:mapping}

In order to study technologically useful or scientifically interesting problems in physics, chemistry, and materials science, it is often necessary to simulate a quantum Hamiltonian. Here we describe the fermionic and bosonic Hamiltonians for which we synthesize a circuit to produce an effective single Trotter step. For each type of particle, we consider a model with a low scaling exponent for the number of Pauli operator terms in the Hamiltonian, what we refer to as a ``low-density'' model, as well a model with a larger scaling exponent, what we refer to as a ``high-density'' model. The Fermi-Hubbard and Bose-Hubbard models are our low-density examples. Our high-density bosonic example is the vibronic problem from chemistry. Finally, our high-density fermionic instance is for the molecular electronic structure problem of polyacetylene (with varying chain length). Fermionic Hamiltonians were produced with the help of OpenFermion \cite{McClean_2020}. For the bosonic degrees of freedom, we considered two different encodings to qubits: standard binary and the Gray code \cite{sawaya19_dlev}. We considered the two cases $d=4,8$ for the level truncation for the bosonic model.

\subsubsection{Fermi-Hubbard model}

The Fermi-Hubbard model \cite{abrams97_fermion} is defined as

\begin{equation}
H_{FH} = -t \sum_{ij\sigma} ( a_{i\sigma}^\dag a_{j\sigma} + a_{j\sigma}^\dag a_{i\sigma} ) + U \sum_i n_{i\uparrow} n_{j\uparrow}
\end{equation}

where $a_{i\sigma}^\dag$ and $a_{i\sigma}$ are respectively fermionic creation and annihilation operators for site $i$ and spin $\sigma \in \{\uparrow,\downarrow\}$, $t$ is the coupling term, and $U$ is the repulsion term. We then used Jordan-Wigner \cite{cao19_chemrev} and Bravyi-Kitaev \cite{bk02,tranter15_bk}methods for mapping fermionic degrees of freedom to qubit degrees of freedom, with even numbers of 2 through 100 sites (for a maximum of 200 qubits), with periodic boundary conditions.

The scaling of terms for this model is $\#(H_{FH}) \sim \mc O(N)$ where $N$ is the number of qubits. 

\subsubsection{Polyacetylene}

In second quantized form, the electronic structure Hamiltonian \cite{helgaaker_book,Whitfield2011,McClean2014_jpcl,babbush15_pra} can be written:
\begin{equation}\label{eqref:ham_mol}
H_{PAc} = \sum_{ij} h_{ij} a_i^\dagger a_p + \sum_{ijpq} h_{ijpq} a_i^\dagger a_j^\dagger a_p a_q
\end{equation}
where spin-orbitals are labeled by $\{i,j,p,q\}$, $h_{ij}$ and $h_{ijpq}$ are one- and two-electron integrals, and an arbitrary number of couplings may be present.

In order to gain insight into scaling of the ultra-greedy search algorithm for molecular electronic structure, we prepared Hamiltonians for a simplified model of cis-polyacetylene. For 2 through 20 carbon atoms, we prepared molecules $C_{n_C}H_{n_C+2}$ for even numbers of carbon atoms $n_C$, and the $C_{n_C}H_{n_C+2}^-$ anion for odd $n_C$. Using anions for odd $n_C$ ensures that none of the molecules have radical electrons, which would lead to a significant qualitative change on the molecular properties. Approximate geometries were found using the software Avogardo 1.2 \cite{avogadro2012} by optimizing with the universal force field (UFF) \cite{rappe92_uff}. We used the minimal STO-3G basis set to perform Hartree-Fock calculations in Psi4 \cite{psi4_2017} and OpenFermion2Psi4 \cite{McClean_2020}. The active space was chosen by filling the lowest $4n_C$ spin-orbitals in the canonical orbital basis, and removing the $4n_C$ highest-energy spin-orbitals, leaving an active space of $4(n_c+1)$ spin-orbitals. All terms smaller than $10^{-6}$ Hartrees in the second quantized Hamiltonian were removed. The Jordan-Wigner and Bravyi-Kitaev transformations were performed. The purpose of this data set is to investigate scaling for the Trotterization algorithm on molecular-related circuits---for truly accurate calculations, different choices would be made regarding geometry optimization, basis size, active space selection, and truncation of small terms.

The scaling of terms for this model is $\#(H_{PAc}) \sim \mc O(N^4)$, where $N$ is the number of qubits used to encode the spin-orbitals. 

\subsubsection{Bose-Hubbard model}

The Bose-Hubbard model \cite{fisher89,bloch_review} is defined as

\begin{equation}
H_{BH} = -t \sum_{ij}^N h_{ij} b_i^\dagger b_j + U \sum_{i}^N b_i^\dagger b_i (b_i^\dagger b_i - 1)
\end{equation}

where $t$ is a coupling term, $U$ is the single site interaction term, and $b_i^\dagger$ and $b_i$ are respectively bosonic creation and annihilation operators for particle $i$. We used bosonic cutoffs $d=4$ and 8, as well as the two bosonic encodings, standard binary( std) and Gray code (gray), and even site numbers $N$ from 2 through 100.

The scaling of terms for this model is $\#(H_{BH}) \sim \mc O(N)$ where $N$ is the number of qubits.

\subsubsection{Vibronic Hamiltonian}

Vibronic (vibrational + electronic) transitions are ubiquitous in chemistry, occurring for example during absorption of light. Because distinct electronic potential energy surfaces (PESs) do not yield identical vibrational normal modes, there is a mixing of modes during the transition between surfaces, with each transition having a distinct intensity (known as a Franck-Condon factor). The mixing of normal modes is determined by the unitary Duschinsky mixing matrix, denoted $\mathbf S$, and the frequencies $\omega_{sj}$ of mode $j$ on surface $s$. Here we summarize the final Hamiltonian used in this work; more details for computations of this class can be found in  \cite{sawaya19_vibronic,mcardle19_qvibr, sawaya20_vibrspec, ollitrault20_reiher_qvibr}.
The first step is to express the vibrational modes of surface $B$ in the normal mode basis of surface $A$. This is done with the transformations

\begin{equation}
\vec q_B = [q_{B0},q_{B1},q_{B2},\cdots]^T = \mathbf{\Omega_B S \Omega_A^{-1}} \vec q_A + \vec\delta
\end{equation}
\begin{equation}
\vec p_B = [p_{B0},p_{B1},p_{B2},\cdots]^T =\mathbf{\Omega_B^{-1} S \Omega_A} \vec p_A
\end{equation}

where $\vec q_s$ and $\vec p_s$ are respectively position and momentum operators for surface $s$ and $\mathbf{\Omega}_s = diag( [\omega_{s1},...,\omega_{sM}] )^{\half}$. Note that $\vec q_B$ and $\vec p_B$ are vectors \textit{of operators}, not vectors of scalars. 

The Hamiltonian is then a sum of independent Harmonic oscillators:
\begin{equation}\label{eq:fc}
\hat H_{FC} = \frac{1}{2} \sum_j^M \omega_{Bj} (q_{Bj}^2 + p_{Bj}^2)
\end{equation}

where $M$ is the number of vibrational modes. For implementation in a quantum device, $\vec q_B$ and $\vec p_B$ are expressed in terms of $\vec q_A$ and $\vec p_A$.

We encoded vibronic Hamiltonians of a fully dense $\mathbf S$ where all $\omega_{Ak}$ and $\omega_{Bk}$ are unique. This provides an upper bound to the resource count required for a vibronic problem of $M$ modes; in real molecules, $\mathbf S$ often has sparsity or additional structure that can be exploited. Hence each Hamiltonian is parametrized by the encoding type, the number of modes $M$, and the number of levels ($d$) allowed in the boson. We used two encodings (Gray and std), $M=6$ through 102, and bosonic cutoffs of $d=4$ and 8. 

The scaling of terms for this model is $\#(H_{FC}) \sim \mc O(N^2)$ where $N$ is the number of qubits used to encode the vibrational modes. 

\subsection{Results}

\begin{figure*}
\centering
\begin{tabular}{cc}
\subfloat[Number of TQE gates per term \label{fig:fh_gates}]{\includegraphics[scale=.6]{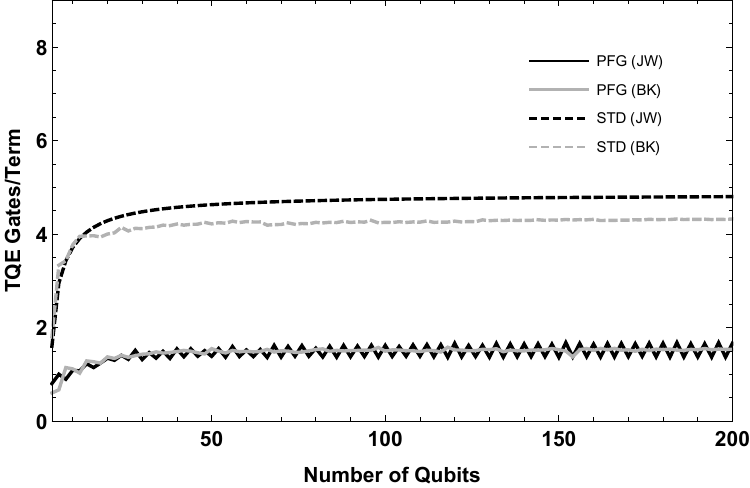}}&
\subfloat[Depth\label{fig:fh_depth}]{\includegraphics[scale=.6]{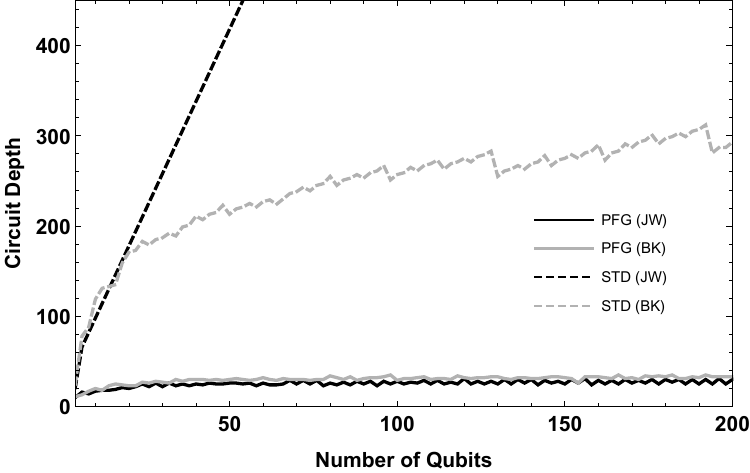}}
\end{tabular}
\caption{Plots for the results of synthesizing an effective Trotter step for the Fermi-Hubbard model.}\label{fig:fh_results}
\end{figure*}

\begin{figure*}
\centering
\begin{tabular}{cc}
\subfloat[Number of TQE gates per term \label{fig:pac_gates}]{\includegraphics[scale=.6]{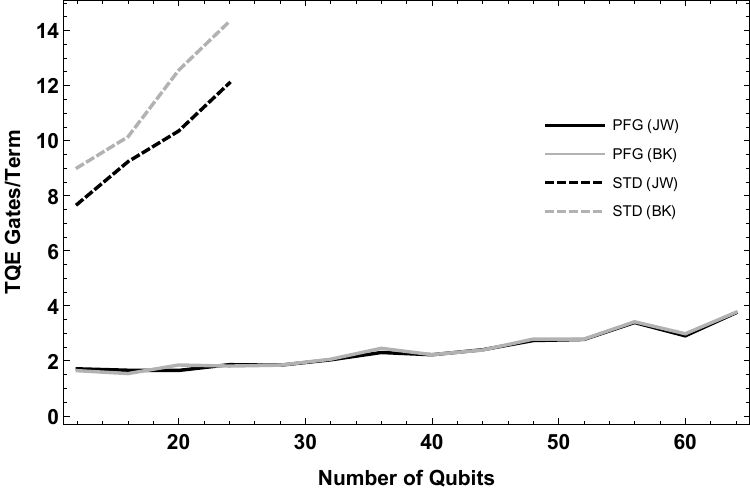}}&
\subfloat[Depth\label{fig:pac_depth}]{\includegraphics[scale=.64]{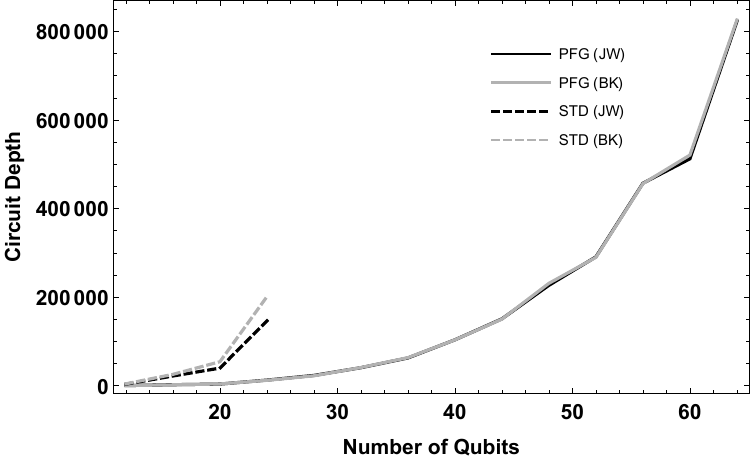}}
\end{tabular}
\caption{Plots for the results of synthesizing an effective Trotter step for the polyacetylene model.}\label{fig:pac_results}
\end{figure*}

\begin{figure*}
\centering
\begin{tabular}{cc}
\subfloat[Number of TQE gates per term; $d=4$ \label{fig:bh4_gates}]{\includegraphics[scale=.58]{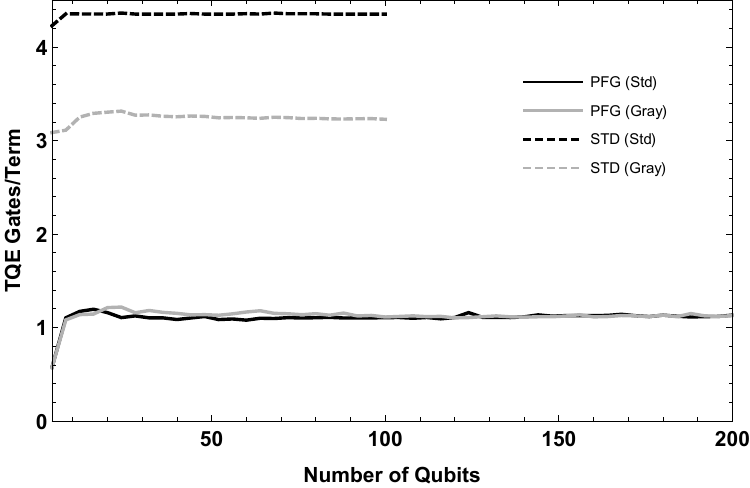}}&
\subfloat[Depth; $d=4$\label{fig:bh4_depth}]{\includegraphics[scale=.64]{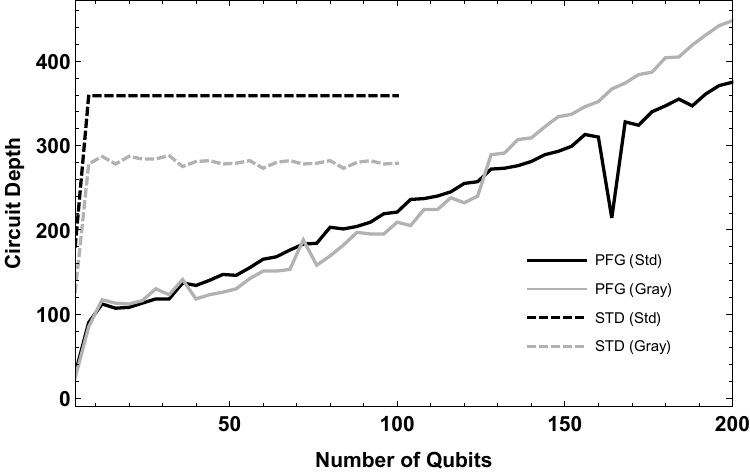}} \\
\subfloat[Number of TQE gates per term; $d=8$ \label{fig:bh8_gates}]{\includegraphics[scale=.58]{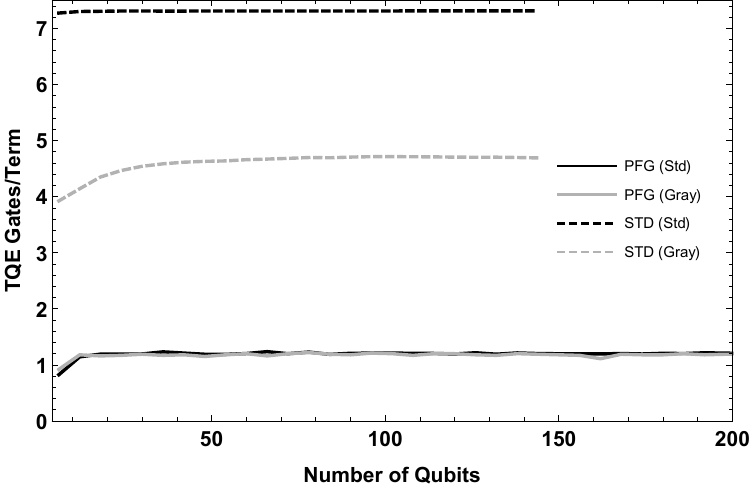}}&
\subfloat[Depth; $d=8$\label{fig:bh8_depth}]{\includegraphics[scale=.64]{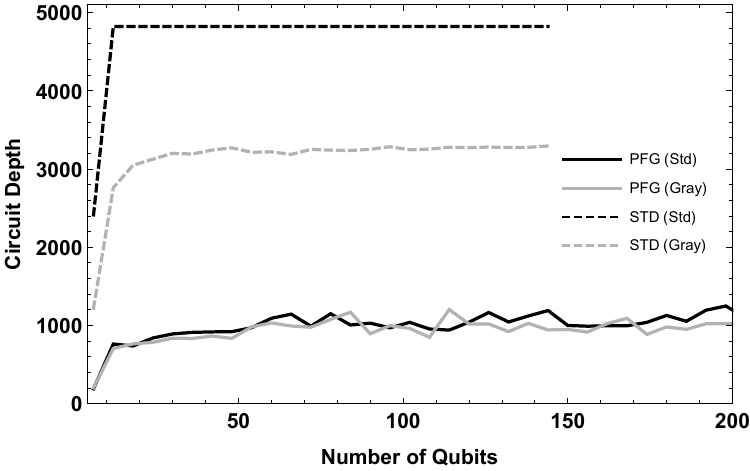}}
\end{tabular}
\caption{Plots for the results of synthesizing an effective Trotter step for the Bose-Hubbard model.}\label{fig:bh_results}
\end{figure*}

\begin{figure*}
\centering
\begin{tabular}{cc}
\subfloat[Number of TQE gates per term; $d=4$ \label{fig:vib4_gates}]{\includegraphics[scale=.58]{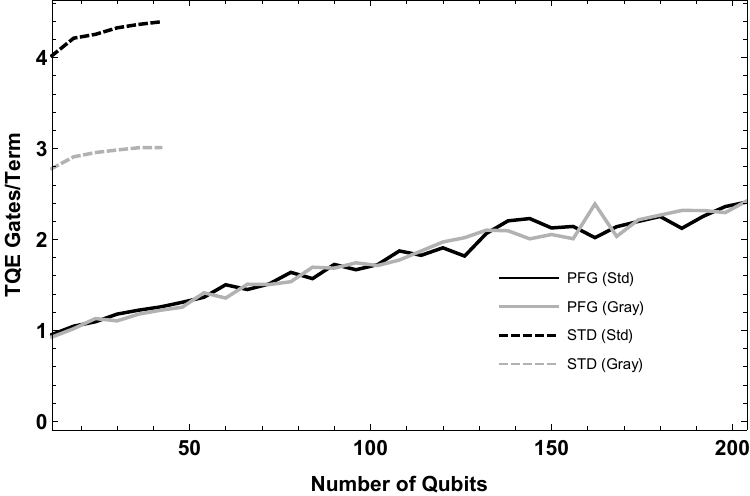}}&
\subfloat[Depth; $d=4$\label{fig:vib4_depth}]{\includegraphics[scale=.58]{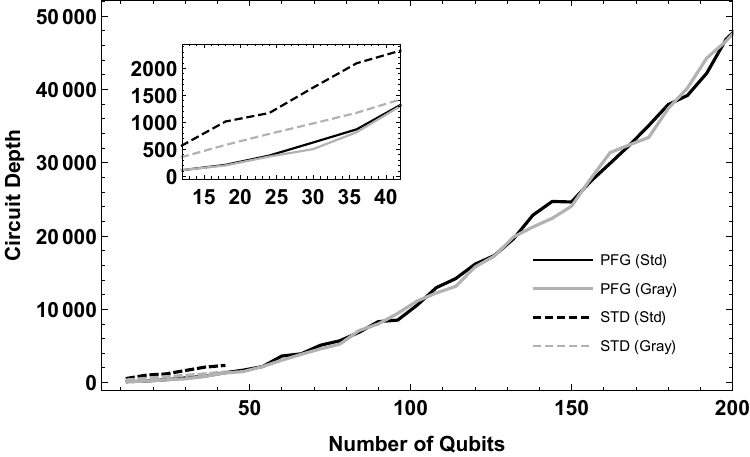}} \\
\subfloat[Number of TQE gates per term; $d=8$ \label{fig:vib8_gates}]{\includegraphics[scale=.58]{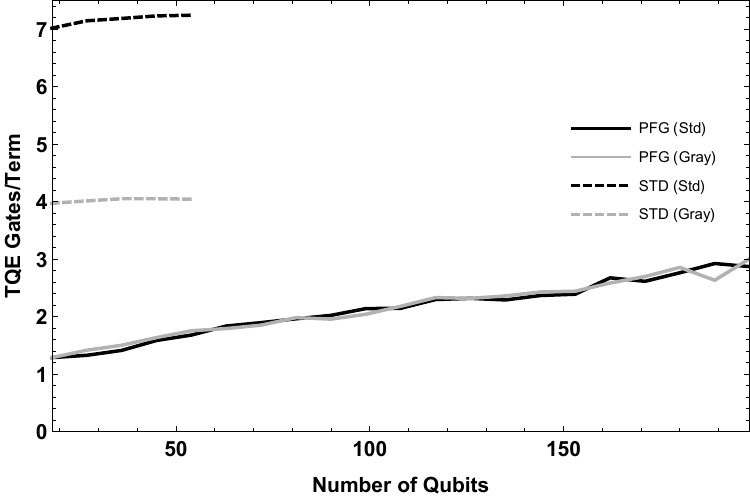}}&
\subfloat[Depth; $d=8$\label{fig:vib8_depth}]{\includegraphics[scale=.58]{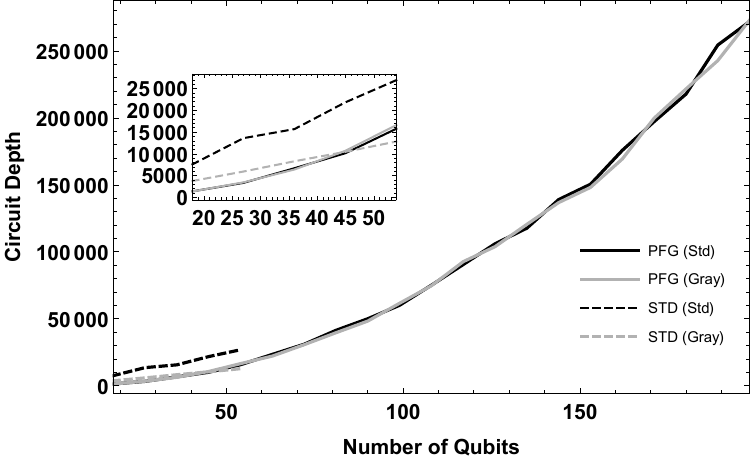}}
\end{tabular}
\caption{Plots for the results of synthesizing an effective Trotter step for the Vibronic model. The insets show a zoomed-in section where both methods produce results. }\label{fig:vib_results}
\end{figure*}

The models described above were synthesized from the PoPR form to a circuit of TQE gates and single-qubit rotations using the ultra-greedy PFG algorithm described above with a parallelization credit of $c=0.1$,
as well as more ``standard'' methods given by decomposing each Pauli rotation using a CX staircase and appropriate single-qubit Clifford gates and applying a Clifford circuit optimization scheme to the resulting circuit.

For the ``standard'' method, in order to reduce circuit depth, the Pauli terms are first placed in sets for which all terms commute within each set. Next, each term is Trotterized using the well-known CNOT staircase construction. The third step is to map the resulting quantum circuit onto a graph, where each gate is a node and edges are present when two gates are adjacent on the same qubit. Finally, we loop through the vertices of this graph, cancelling out adjacent gates where possible. A time limit of 600 seconds was set on this last step, meaning that there may be some remaining gate cancellations that are neglected.

Figs. \ref{fig:fh_results}, \ref{fig:pac_results}, \ref{fig:bh_results} and \ref{fig:vib_results} each plot the resulting average number of TQE gates 
needed per Pauli Hamiltonian term and circuit depth for one effective Trotter step\footnote{This does not include the return path as we assume retracing.} as a function of qubit number.
Due to the parallelization and overall efficiency of the the ultra-greedy PFG algorithm implementation, the PFG methods were able to synthesize much larger examples, in which case insets for some of these plots are a blown-up section of the graph where both methods produced results.

In general, one can see that the PFG method always results in fewer TQE gates per term and in almost every case a lower depth over standard methods, and often significantly so. In the case of the polyacetylene model, the reduction is as much as an order of magnitude in the largest examples where both methods returned a result. We do note that in the case of the Bosonic models, it appears possible in the case of the Bose-Hubbard model and does in the vibronic model that standard methods overtake our PFG methods when considering depth. This may be an indication of the limitations of greedy search. Looking at the circuits for smaller instances, one can see that while in the beginning the circuits are highly dense, rotations generally begin to rarefy causing ever-longer TQE gate chains in order to reach some member of  the dwindling pool of unapplied rotations. This is to be expected of a greedy method and demonstrates the need for more sophisticated PFG methods for finding a Trotter path. Though we do not see it here, it can happen that our methods generate more TQE gates than standard methods, but result in a lower depth. Again this is due in part to the limitations inherent in greedy search, but also attributable to the addition of the parallelization credit in Eq. \eqref{eq:cost}. Such considerations allow the method to add more TQE gates if doing so results in a lower depth. Multiple runs with different values for the parallelization credit were not performed for this study, though it was shown in Ref.~\cite{PaykinSchmitz2023PCOAST} that although a different context, varying the parallelization credit can significantly reduce circuit metrics.

\begin{figure*}
\centering
\begin{tabular}{cc}
\subfloat[Fermi-Hubbard\label{fig:fh_time}]{\includegraphics[scale=.58]{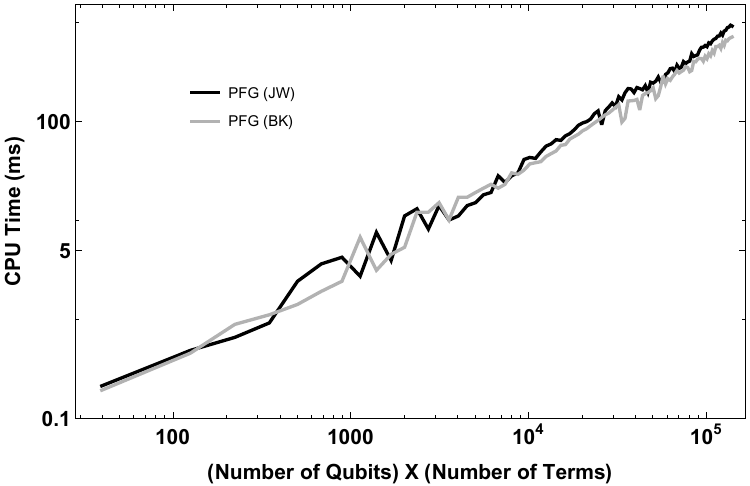}}&
\subfloat[Polyacetylene\label{fig:pac_time}]{\includegraphics[scale=.58]{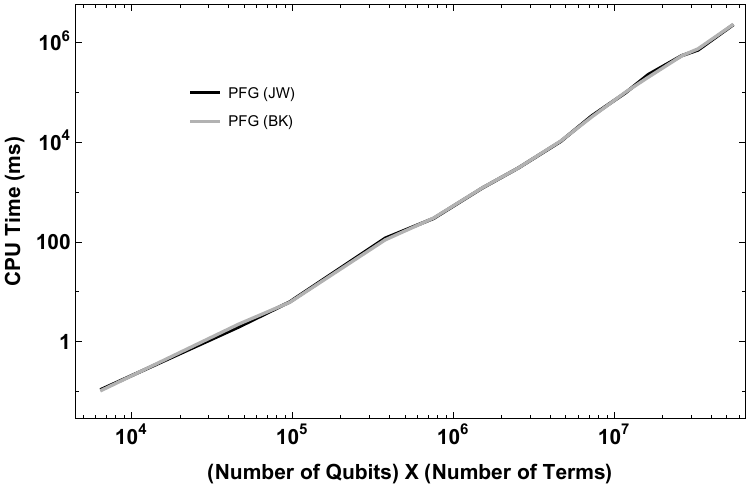}}\\
\subfloat[Bose-Hubbard; $d=4$\label{fig:bh4_time}]{\includegraphics[scale=.58]{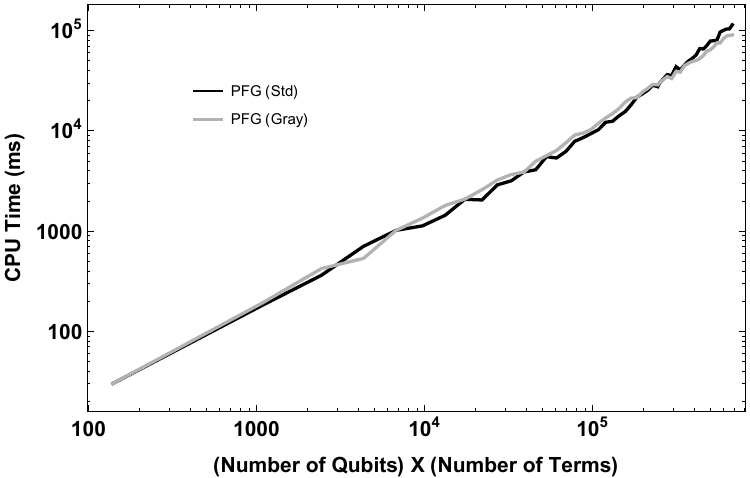}}&
\subfloat[Bose-Hubbard; $d=8$\label{fig:bh8_time}]{\includegraphics[scale=.58]{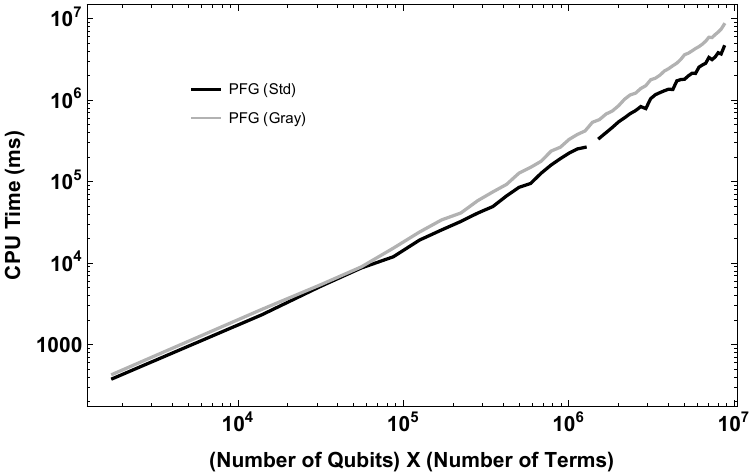}}\\
\subfloat[Vibronic; $d=4$\label{fig:vib4_time}]{\includegraphics[scale=.58]{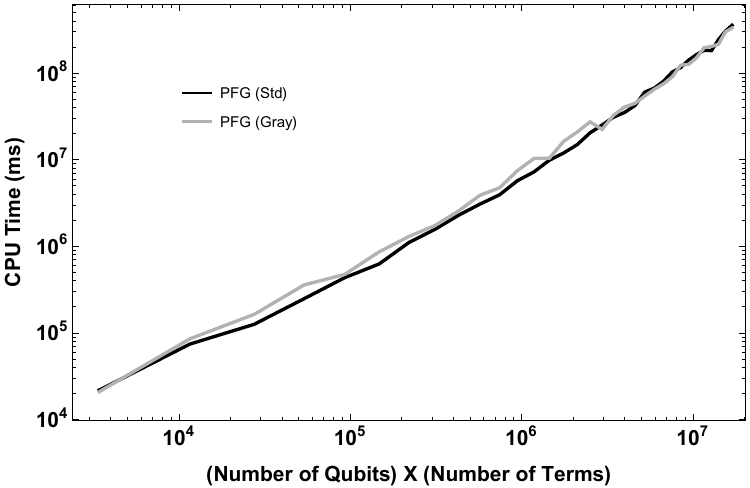}}&
\subfloat[Vibronic; $d=8$\label{fig:vib8_time}]{\includegraphics[scale=.58]{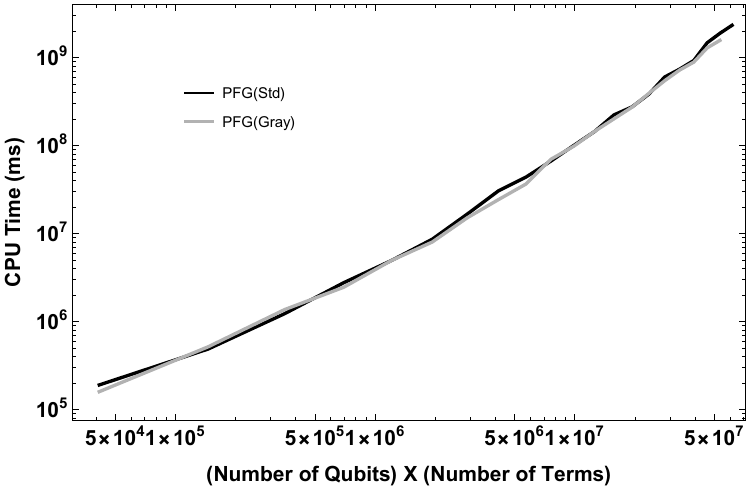}}
\end{tabular}
\caption{Log-log plots of cpu -time (wall-clock time times number of cpu processes used) as a function of number of qubits times number of Pauli Hamiltonian terms. }\label{fig:time}
\end{figure*}

Fig. \ref{fig:time} shows the CPU time to complete the PFG ultra-greedy search as a function of $N \#(H)$ on a log-log plot for each of the models described here.  The PFG methods utilized as many 100 parallel processes for some of the larger problem instances presented in this paper, some of which were multi-threaded processes. In every case we see an expected scaling exponent between 1 and 2. Table \ref{scaling} gives the fitted scaling exponent. Here we see that the scaling exponent does appear to have some dependency on the density of the model. This could be due in part to the use of parallelization for larger models which comes with considerable communication overhead. See Appendix~\ref{ap:parallel} for a description of this. This is also consistent with the fact that the $d=4$ encodings have a generally smaller exponent than $d=8$ for Bosonic models, which is true for both low and high density models. Overall, this study of the time scaling of the algorithm demonstrates the PFG ultra-greedy search algorithm is both efficient and scalable, as also demonstrated by the large problem instances for which the method was able to produce a circuit. 

To be fair to the standard methods, the code used was not optimized or parallelized to the same extent as the PFG ultra-greedy search code. Furthermore, arbitrary timeout limits were set for practical reasons due to the large data set studied. Thus we do not attempt to compare time scaling between the two methods.

\begin{figure}
\centering
\begin{tabular}{|c|c|c|}
\hline
Model & $\alpha$(JW/Std) &$\alpha$( BK/Gray)\\
\hline
Fermi-Hubbard & 1.128 & 1.180\\
Polyacetylene & 1.672 & 1.814\\
Bose-Hubbard $d = 4$ & 1.324 & 1.122\\
BoseHubbard $d=8$ & 1.437 & 1.523\\
Vibronic $d=4$ & 1.583 & 1.524\\
Vibronic $d=8$ & 1.846 & 1.698\\
\hline
\end{tabular}
\caption{Table of scaling exponent for the time scaling of the PFG ultra-greedy search algorithm as a function of number of qubits times number of Pauli Hamiltonian terms, i.e.  (CPU time to complete search) $\sim (N \#(H))^\alpha$. } \label{scaling}
\end{figure}

\section{Outlook and Conclusions}\label{sec:conclusion}

In this paper, we have motivated and described the theory behind the Pauli Frame Graph and how it provides a perspective on synthesizing circuits from the product-of-Pauli rotations form. From there, we focused on the use-case which synthesizes a Trotter step in the Trotter-Suzuki decomposition for the simulation of a Hamiltonian often used for applications in physics, material science, and chemistry as well as subroutines in many common quantum algorithms. Using this perspective, we also developed the ultra-greedy search algorithm for synthesizing an efficient Trotter step, especially when using the retrace method. We then demonstrated the power of the algorithm by applying the method to four models, two of which were Fermionic, two Bosonic, two low density and two high density. In nearly every case, the results of our algorithm produced circuits with significantly lower depth over the more conventional method and fewer two-qubit entangling gates. We also showed that this algorithm is both scalable and efficient. 

Though we have found a reasonable circuit synthesis algorithm here, the real power of this work is in the paths it opens for future research. These methods have already been refined and adapted for general circuit synthesis \cite{PaykinSchmitz2023PCOAST} and measurement reduction for a mutually-commuting set of Pauli operators\cite{schmitz2023PCOAST} as a part of the Pauli-based Cricuit Optimization, Analysis and Synthesis Toolchain (PCOAST). This includes future work in PCOAST to adapt all such use cases to include limited qubit connectivity. Moreover, the ultra-greedy methodology is the simplest, least sophisticated method for using the PFG perspective of circuit synthesis. Thus we look to develop more sophisticated synthesis methods in the future which balance both circuit optimization as well as overall unitary accuracy, such that we reduce the total circuit cost for the entire Trotterized simulation. Finally, we look to add this functionality to be used by the  Intel$^{\text{\textregistered}}$ Quantum SDK \cite{Khalate2022} with extensions for algorithm abstraction \cite{schmitz2023functional}  such as Hamiltonian simulation.

\begin{acknowledgments}

This research used resources of the National Energy Research Scientific Computing Center (NERSC), a U.S. Department of Energy Office of Science User Facility located at Lawrence Berkeley National Laboratory, operated under Contract No. DE-AC02-05CH11231 using NERSC award DDR-ERCAP0018781.

\end{acknowledgments}

\appendix

\section{Proof of Free and Transitive Group Properties for the Symplectic Group Acting on Pauli Frames}\label{ap:group}

In this section, we prove that the group action of the symplectic group on Pauli frames is free and transitive. See main text for definitions.

\begin{proof}

To prove transitivity, we must show that for frames $B_1= \{(s_i, \tilde s_i)\}_{0\leq i<N}, B_2=\{( t_i, \tilde t_i)\}_{0\leq i<N}$, there exist a symplectic automorphism $\left(\gamma: \mc P \to \mc P\right) \in \Sp(2N, \mb F_2)$ such that $\gamma(B_1)=B_2$. We define such a map as,
\begin{align}
p \mapsto \sum_i \left( \lambda (p, s_i) \tilde t_i + \lambda(p, \tilde s_i) t_i \right). 
\end{align}

Clearly this is linear. Also, for any $p,q \in \mc P$ we have that
\begin{align}
\lambda(\gamma(p),\gamma(q)) =& \sum_i \big( \lambda(p, s_i) \lambda(\tilde t_i, \gamma(q))  \nonumber \\
 &+ \lambda(p, \tilde s_i) \lambda(t_i, \gamma(q)) \big) \nonumber \\
=& \sum_i \left( \lambda(p, s_i) \lambda(\tilde s_i, q) + \lambda(p, \tilde s_i) \lambda(s_i, q) \right) \nonumber \\
=& \lambda(p, q).
\end{align}
Thus $\gamma \in \Sp(2N, \mb F_2)$. Finally, if we apply $\gamma$ to any member of $B_1$, we have $\gamma(s_i) = t_i$ and $\gamma(\tilde s_i) = \tilde t_i$ for all $i$. Therefore, $\gamma(B_1) = B_2$ and our group action is transitive.

To show that this group action is free, we must show that if for some  $\gamma, \delta \in \Sp(2N, \mb F_2)$ there exists a $B$ such that $\gamma(B)= \delta(B)$, then $\gamma=\delta$. So suppose we do have such a Pauli frame for some $\gamma$ and $\delta$. As $\gamma(B)= \delta(B)$ is a basis for $\mc P$, by linearity for every $p \in \mc P$, 
\begin{align}
\gamma(p) =& \sum_i \left( \lambda(p, s_i) \gamma(\tilde s_i) + \lambda(p, \tilde s_i) \gamma(s_i)\right) \nonumber \\
=& \sum_i \left( \lambda(p, s_i) \delta(\tilde s_i) + \lambda(p, \tilde s_i) \delta(s_i)\right) \nonumber \\
=& \delta(p).
\end{align}
 Therefore, $\gamma =\delta$ and our group action is free. 

\end{proof}

\section{Deriving the Action and Signs for Clifford Gates}\label{ap:signs}
In this Section, we discuss the general methods for determining the transformation and sign of Pauli operators and Pauli frame elements under the action of two-qubit Clifford gates. We do this by first defining two general Pauli-based Clifford gates. For Hermitian $p\in \mc P'$ we define the {\it root-of} gate
\begin{align}
\sqrt p = \exp\left(- i\frac{\pi}{4} p\right).
\end{align}
This generalizes the phase gate for any given Pauli operator, i.e $\p= \sqrt{Z}$. It should be clear that the conjugation rule for any other $q \in \mc P'$ is
\begin{align}\label{conj1}
\sqrt p q \sqrt p^\dagger =
\begin{cases}
p &  \lambda(p,q) =0 \\
 -i pq & \lambda(p,q)=1
 \end{cases}.
\end{align}
For two Hermitian, mutually commuting Pauli operators $p, q \in \mc P'$, we define the {\it AND-Sign} (ASign) gate
\begin{align}
\text{ASign}(p, q) =& \exp\left(-i \frac{\pi}{4}(1-p)(1-q) \right) \nonumber \\
 \propto& \sqrt{p}^\dagger \sqrt{pq} \sqrt{q}^\dagger.
\end{align}
This gate applies a minus sign to eigenstates for which $p$ and $q$ evaluate to -1. This generalizes the TQE gates, i.e. $CX_{01} = \text{ASign}(Z_0, X_1)$. Using Eq. \eqref{conj1}, we can see the conjugation rule for any $r \in \mc P'$ is
\begin{align} \label{conj2}
&\text{ASign}(p,q) r \text{ASign}(p,q) = \nonumber \\
& 
\begin{cases}
r & \lambda(p,r) = \lambda(q, r) = 0\\
qr & \lambda(p,r) =1, \lambda(q,r) =0\\
pr & \lambda(p,r) =0, \lambda(q,r) = 1\\
-pqr & \lambda(p,r) = \lambda(q,r) = 1
\end{cases}.
\end{align}

To derive the gate action on Pauli operators and Pauli frames, one recognizes that all two qubit gate action, both forward and backward, can be reduced to these conjugation rules, if one implements sign/phase-aware Pauli multiplication. 

In the case of forward action for two-qubit entangling (TQE) gates, transformation of the Pauli frame is just conjugating each Pauli operator entry. For TQE gate $g_{ij} = \text{ASign}((\sigma_1)_i, (\sigma_2)_j)$ and $\sigma_{1(2)} \in \{X, Y, Z\}$, \ we compute $q = g_{ij} p g_{ij}$ as:
\begin{enumerate}
 \item Let $q\gets p$.
 \item compare the local support of $p$ on qubit $i$ to $\sigma_1$ and if they locally anti-commute, $q \gets (\sigma_2)_j q$.
 \item Repeats step 1 swapping the roles of the qubits.
 \item  include an additional sign if $q$ was modified in both steps 2 and 3. 
 \end{enumerate}
 Note the order of the qubits in this process is irrelevant due to commutation of the relevant operators as can be seen in Eq. \eqref{conj2}. From this, one can derive the resulting gate action, including sign, for all $144 =9 \times 4 \times 4$ possible local interactions. For completeness, we also note the single-qubit multiplication rule
 \begin{align}
 XYZ= i,
 \end{align}
 from which all single qubit multiplication, with sign/phase, can be derived.
 
 Note this discussion implies the assertion made in the main text that for any Pauli operator with support on qubits $i,j$, there exist four TQE gates acting on $i, j$, which reduce its support by one. To see this, suppose the Pauli has support $X$ on $i$ and $Y$ on $j$. To remove support on $i$, we need the TQE gate to have type X on $i$ and not have type Y on $j$. Similarly to reduce support on $j$ we need the TQE gate to have type Y on $j$ and not have type X on $i$. Thus the four gates would be $AX_{ij}, AZ_{ij}, BY_{ij}$ and $CY_{ij}$. This can be extended to any support.  
 
 For backward action by definition, we use Eq. \eqref{conj2} to consider the action of the TQE gate on the origin frame and then swap out the single-qubit Paulis for the elements of the frame. But a TQE gate can only interact with single-qubit Pauli operators for the two qubits on which it acts.  For TQE gate $g_{ij} = \text{ASign}((\sigma_1)_i, (\sigma_2)_j)$ and $\sigma_{1(2)} \in \{X, Y, Z\}$ and $B = \{(s_i, \tilde s_i)\}_{0\leq i<N}$, we compute $B' = g_{ij} B g_{ij}$ 
 \begin{enumerate}
 \item $B' \gets \{(s_i, \tilde s_i)\}_{0\leq i<N}$
 \item Let 
 \begin{align}
 p = \begin{cases}
  \tilde s_j & \text{if } \sigma_2 =  X\\
  -i s_j \tilde s_j & \text{if } \sigma_2 =  Y \\
  \tilde s_j& \text{if } \sigma_2 =  Z
  \end{cases}
  \end{align}
\item For elements in $B'$,
\begin{align}
\begin{cases}
   s_i \gets  s_i p& \text{if } \sigma_1 =  X\\
  s_i \gets s_i p \text{, }\tilde s_i \gets \tilde s_i p& \text{if } \sigma_1 =  Y \\
  \tilde s_i \gets \tilde s_i p& \text{if } \sigma_1 =  Z
  \end{cases}
\end{align}
\item Repeat steps 1 and 2 swapping the role of the qubits.
 \end{enumerate}
Note we use multiplication, not ``binary addition'' because this includes sign and thus order matters for the resulting sign. Also note the extra sign from Eq. \eqref{conj2} does not enter here, so the sign always comes from the multiplication.

\section{Proof of Proposition 1}\label{ap:prop1}

In this Section, we prove the claims of Proposition \ref{prop1} of the main text.

\begin{proof}
 We prove each claim in turn.
\begin{enumerate}
 \item $\supp(p,B) \geq 0$ is trivially true. Furthermore, suppose $\supp(p,B)=0$ for some $p \in \mc P$ and $B\in \mc B$. This is true if and only if $\lambda(s_i,p) \vee \lambda(\tilde s_i, p)=0$ for all $i$ which is true if and only if $\lambda(s_i,p) = \lambda(\tilde s_i, p)=0$ for all $i$. As $B$ is a basis for $\mc P$, the former is true if and only if $p=I$.
  
\item Consider the following differenc,e
\begin{align}\label{eq:pdiff}
\supp(p,B)& + \supp(q,B)- \supp(p+q ,B)\nonumber \\
 = \sum _i& \big( \lambda(s_i,p) \vee \lambda(\tilde s_i, p)+ \lambda(s_i,q) \vee \lambda(\tilde s_i, q) \nonumber \\
 &- \lambda(s_i,p+q) \vee \lambda(\tilde s_i, p+q)\big).
\end{align}
So for each $i$ term we have the general expression $x \vee \tilde x + y \vee \tilde y - (x \oplus \tilde x) \vee (y \oplus \tilde y)$ using the binary linearity of $\lambda$. Such an expression is less than zero if and only if $x \vee \tilde x= y \vee \tilde y=0$ and $(x \oplus \tilde x) \vee (y \oplus \tilde y)=1$ but this is a contradiction as $x \vee \tilde x= y \vee \tilde y=0$ implies that $x = \tilde x =y = \tilde y=0$, in which case$ (x \oplus \tilde x) \vee (y \oplus \tilde y)=0$. Thus every term on the RHS of  Eq. \eqref{eq:pdiff} is greater than zero, implying the LHS must be greater than zero. Therefore, condition (1) is true.

\item To show this is true for all $\cancel E_N$, it is sufficient to show it is true for all its generators. First consider $\h_j$ for some $j$:
\begin{align}
\supp(p, \h_j(B)) =& \sum_{i\neq j}\lambda(s_i,p) \vee \lambda(\tilde s_i, p) \nonumber \\
 &+ \lambda(\tilde s_j,p) \vee \lambda( s_j, p) \nonumber \\
 =& \sum_{i}\lambda(s_i,p) \vee \lambda(\tilde s_i, p) \nonumber \\
 =& \supp(p, B),
\end{align}
where we used the symmetry property of $\vee$. Also consider that $(x\oplus y) \vee y = x \vee y$. So
 
\begin{align}
\supp(p, \p_j(B)) =&  \sum_{i\neq j}\lambda(s_i,p) \vee \lambda(\tilde s_i, p) \nonumber \\
 &+ \lambda(s_j +\tilde s_i,p) \vee \lambda( \tilde s_i, p) \nonumber \\
 =& \sum_{i}\lambda(s_i,p) \vee \lambda(\tilde s_i, p) \nonumber \\
 =& \supp(p, B),
\end{align}
using the binary linearity of $\lambda$ and the above Boolean equivalence. Finally, any qubit permutations such as SWAP trivially satisfies (3) as it only rearrange the terms defining $\supp$. Therefore, (3) is true.

\item First, we show that some member of $B' \in [B]$ can be transformed so as to contain $p$ using $ \supp(p, B)-1$ CXs, and then argue this is the minimum number. So let $d = \supp(p, B)$. This implies there are $d$ qubit indices, $i$ such that $ \lambda(s_i,p) \vee \lambda(\tilde s_i, p)=1$. Let $D$ be the set of these indices. For each such $i \in D$, if $\lambda(s_i,p) = \lambda(\tilde s_i, p)=1$, apply $\h_i \circ \p_i (B)$ for which the $i^{th}$ part of the frame is then $( s_i + \tilde s_i,  s_i)$. If $\lambda(s_i,p) =1, \lambda(\tilde s_i,p)=0$, apply $\h_i(B)$, for which the $i^{th}$ part of this frame is then $(\tilde s_i, s_i)$. All other cases, do nothing. In the resulting frame, which we refer to as $B'$, $\lambda(s_i', p)= 0, \lambda(\tilde s_i',p)=1$ for all $i$, and since $B \to B'$ uses only members of $ \cancel E_N$, $B'\in [B]$. Because $B'$ is a frame, we can use the expansion of $p$ in this frame such that that $p =\sum_{i \in D} s_i'$. One can use $|D|-1$ CNOTs to reduce this sum and reach a frame $B'' \ni p$. Furthermore, because this is the unique expansion of $p$ in terms of this frame, this is the minimum number of CXs needed to reach such a frame.
\end{enumerate}
\end{proof}

\section{Detailed Description of the Ultra-greedy PFG Search Algorithm} \label{ap:algo}

In this Section we provide a detailed description of the ultra-greedy PFG search algorithm  for a Hamiltonian path. For what follows, $\text{bin}(p)$ converts $p$ to its signed binary representation in the origin frame and $N$ qubits. Recall $c$ isa free parameter of the cost function $cost_c(g)\leq1$ as discussed in the main text.

\begin{algorithm}[H]
\caption{PFG Ultra-greedy Search} \label{alg2}
\hspace*{2em} \textbf{Input:} List of Hamiltonian terms and angles $H=\{(\theta_\alpha, p_\alpha)\}$ \\
\hspace*{2em} \textbf{Return:} List of circuit elements $C_\text{return}$ 
\begin{algorithmic}[1]
\State $C_\text{return}\gets\{\}$ i.e. is an empty circuit. 
\State$ham \gets \{\text{bin}(p_\alpha)\}$.
\While{$ham\neq \emptyset$}
\State $minsup \gets N$
\State $ham_{min}=\{\}$
\State $cost= 1.1$
\State $g_{min} \gets I$
\For{$p \in ham$}
\If{\supp($p$) =1}
\State Add rotation to $C_\text{return}$ according to local support for angle $\pm\theta_\alpha$ with sign given by sign of $p$.
\State Remove $p$ from $ham$. 
\ElsIf {$\supp(p)= minsup$}
\State Add $p$ to $ham_{min}$
\ElsIf{$2\leq \supp(p)<minsup$}
\State Clear $ham_{min}$ and add $p$ 
\State $minsup \gets \supp(p)$
\EndIf
\EndFor
\For{$p \in ham_{min}$}
\For{all gates $g$ which reduces support of $p$}
\If{$cost_{c}(g) < cost$}
\State $g_{min}\gets g$.
\State $cost \gets cost_{c}(g)$
\EndIf
\EndFor
\EndFor
\State Add $g_{min}$ to $C_\text{return}$.
\State Update all members of $ham$ using the TQE transformation rules for $g_{min}$.
\EndWhile
\end{algorithmic}
\end{algorithm}

\section{Details of Parallelization of the Ultra-Greedy Search Algorithm}\label{ap:parallel}

In this Section, we discuss the methods used for parallelizing the ultra-greedy search algorithm described in Section \ref{ap:algo} and the main text. The code used in this study leveraged MPI to implement the parallelized program described here. It was found that the best method for parallelizing the algorithm was to assign each process (core or thread) a set of Hamiltonian terms for which that process is responsible. This does requires a considerable amount of communication between the processes. The parallel program proceeded as follows:
\begin{itemize}
\item Any immediate support one terms are converted to a rotation and added to the output cricuit by the rank 0 process. (see footnote below for reason.)
\item Each remaining term is assigned and  communicated to its owning process.
\item Each process has a local copy of the leading time edge for each qubit, what we deem the {\it schedule}, which is initially communicated to each process.
\item The total global number of remaining terms is held locally and communicated to each process. This value being greater than zero is used as the continue condition for the main loop of the program.
\item each process enters the main loop for which the following happens:
\begin{itemize} 
\item Each process searches its owned terms for those with support one. Such terms are converted to a rotation and added to a local circuit list with a timing stamp as obtained from the schedule.
\item Once each process completes the above search, the schedule is synced between the processes by finding and sharing the global maximum value for each qubit.\footnote{ Note this method relies on the fact that no process will try to schedule a rotation on the same qubit as any another. This is guaranteed by two facts: 1) no Pauli terms are duplicated, and thus each process has a distinct list of Pauli operator terms. 2) each pass through the main loop only applies one TQE gate. It should be clear that the application of a TQE gate only provides two new possible rotations not available in the previous frame. Furthermore, these are alway on distinct qubits. Together, one can see that aside for the initial case (as was handled at the beginning by the rank 0 process), each pass through the main loop can not result in two rotation gates being scheduled on the same qubit. \label{fn1} }
\item  Simultaneously with steps above, each process searches for its minimum support terms. The resulting minimum cost is shared with the rank 0 process and a global minimum is communicated to every process. If the process has a local cost greater than the global cost, its local list of minimum support terms is cleared.
\item Each process collects all TQE gates which would reduce its local list of minimum Pauli operator terms. These lists are collected by the master, duplicates are removed, and the final global list of TQE gates is communicated to each process.
\item For each candidate TQE gate, the unnormalized average change in support is calculated locally for the terms owned by that process, and the global sum collected by the rank 0 process. The rank 0 process then normalizes this value by the global number of terms and completes the cost function calculation by including the parallelization credit.
\item Once all candidate TQE gates are evaluated, the rank 0 process holds the minimum cost gate. This gate is communicated to the other processes. The master adds this gate to its circuit list with a time stamp from the schedule (maximum time between the two qubits acted upon).
\item Each process updates all of its owned terms with the new TQE gate action, and its local schedule. 
\item The total global number of terms remaining is found and communicated to all processes.
\end{itemize}
\item Once all processes exit the main loop, each process communicates to the rank 0 process its list of circuit elements, this list is combined into one list and the list is sorted in time order.
\item The rank 0 process returns this final circuit list.
\end{itemize}

Another method for parallelization we investigated was to have every process work on the full list of terms, while splitting the candidate TQE gate cost evaluations evenly between the processes. Though this required far less communication between the processes, it was found that the speed-up was limited by the search for the minimum support Pauli terms as this step could not be parallelized other than to do the above. So although the above method of splitting terms between the processes was considerably more complex, it resulted in a parallelized program which more closely matched the ideal speed-up. Furthermore, other non-parallelization methods exist for reducing the cost of gate cost evaluation, but were not exploited in the study of the main text. We hope to explore additional speed-ups in future work.

\bibliography{main}

\begin{thebibliography}{77}%
\makeatletter
\providecommand \@ifxundefined [1]{%
 \@ifx{#1\undefined}
}%
\providecommand \@ifnum [1]{%
 \ifnum #1\expandafter \@firstoftwo
 \else \expandafter \@secondoftwo
 \fi
}%
\providecommand \@ifx [1]{%
 \ifx #1\expandafter \@firstoftwo
 \else \expandafter \@secondoftwo
 \fi
}%
\providecommand \natexlab [1]{#1}%
\providecommand \enquote  [1]{``#1''}%
\providecommand \bibnamefont  [1]{#1}%
\providecommand \bibfnamefont [1]{#1}%
\providecommand \citenamefont [1]{#1}%
\providecommand \href@noop [0]{\@secondoftwo}%
\providecommand \href [0]{\begingroup \@sanitize@url \@href}%
\providecommand \@href[1]{\@@startlink{#1}\@@href}%
\providecommand \@@href[1]{\endgroup#1\@@endlink}%
\providecommand \@sanitize@url [0]{\catcode `\\12\catcode `\$12\catcode
  `\&12\catcode `\#12\catcode `\^12\catcode `\_12\catcode `\%12\relax}%
\providecommand \@@startlink[1]{}%
\providecommand \@@endlink[0]{}%
\providecommand \url  [0]{\begingroup\@sanitize@url \@url }%
\providecommand \@url [1]{\endgroup\@href {#1}{\urlprefix }}%
\providecommand \urlprefix  [0]{URL }%
\providecommand \Eprint [0]{\href }%
\providecommand \doibase [0]{https://doi.org/}%
\providecommand \selectlanguage [0]{\@gobble}%
\providecommand \bibinfo  [0]{\@secondoftwo}%
\providecommand \bibfield  [0]{\@secondoftwo}%
\providecommand \translation [1]{[#1]}%
\providecommand \BibitemOpen [0]{}%
\providecommand \bibitemStop [0]{}%
\providecommand \bibitemNoStop [0]{.\EOS\space}%
\providecommand \EOS [0]{\spacefactor3000\relax}%
\providecommand \BibitemShut  [1]{\csname bibitem#1\endcsname}%
\let\auto@bib@innerbib\@empty
\bibitem [{\citenamefont {JavadiAbhari}\ \emph {et~al.}(2015)\citenamefont
  {JavadiAbhari}, \citenamefont {Patil}, \citenamefont {Kudrow}, \citenamefont
  {Heckey}, \citenamefont {Lvov}, \citenamefont {Chong},\ and\ \citenamefont
  {Martonosi}}]{scaffcc2015}%
  \BibitemOpen
  \bibfield  {author} {\bibinfo {author} {\bibfnamefont {A.}~\bibnamefont
  {JavadiAbhari}}, \bibinfo {author} {\bibfnamefont {S.}~\bibnamefont {Patil}},
  \bibinfo {author} {\bibfnamefont {D.}~\bibnamefont {Kudrow}}, \bibinfo
  {author} {\bibfnamefont {J.}~\bibnamefont {Heckey}}, \bibinfo {author}
  {\bibfnamefont {A.}~\bibnamefont {Lvov}}, \bibinfo {author} {\bibfnamefont
  {F.~T.}\ \bibnamefont {Chong}},\ and\ \bibinfo {author} {\bibfnamefont
  {M.}~\bibnamefont {Martonosi}},\ }\href
  {https://doi.org/https://doi.org/10.1016/j.parco.2014.12.001} {\bibfield
  {journal} {\bibinfo  {journal} {Parallel Computing}\ }\textbf {\bibinfo
  {volume} {45}},\ \bibinfo {pages} {2} (\bibinfo {year} {2015})},\ \bibinfo
  {note} {computing Frontiers 2014: Best Papers}\BibitemShut {NoStop}%
\bibitem [{\citenamefont {Steiger}\ \emph {et~al.}(2018)\citenamefont
  {Steiger}, \citenamefont {H{\"{a}}ner},\ and\ \citenamefont
  {Troyer}}]{projQ2018}%
  \BibitemOpen
  \bibfield  {author} {\bibinfo {author} {\bibfnamefont {D.~S.}\ \bibnamefont
  {Steiger}}, \bibinfo {author} {\bibfnamefont {T.}~\bibnamefont
  {H{\"{a}}ner}},\ and\ \bibinfo {author} {\bibfnamefont {M.}~\bibnamefont
  {Troyer}},\ }\href {https://doi.org/10.22331/q-2018-01-31-49} {\bibfield
  {journal} {\bibinfo  {journal} {{Quantum}}\ }\textbf {\bibinfo {volume}
  {2}},\ \bibinfo {pages} {49} (\bibinfo {year} {2018})}\BibitemShut {NoStop}%
\bibitem [{\citenamefont {Koch}\ \emph {et~al.}(2019)\citenamefont {Koch},
  \citenamefont {Wessing},\ and\ \citenamefont {Alsing}}]{qiskit2019}%
  \BibitemOpen
  \bibfield  {author} {\bibinfo {author} {\bibfnamefont {D.}~\bibnamefont
  {Koch}}, \bibinfo {author} {\bibfnamefont {L.}~\bibnamefont {Wessing}},\ and\
  \bibinfo {author} {\bibfnamefont {P.~M.}\ \bibnamefont {Alsing}},\
  }\href@noop {} {\bibinfo {title} {Introduction to coding quantum algorithms:
  A tutorial series using qiskit}} (\bibinfo {year} {2019}),\ \Eprint
  {https://arxiv.org/abs/arXiv:1903.04359} {arXiv:1903.04359} \BibitemShut
  {NoStop}%
\bibitem [{\citenamefont {Sivarajah}\ \emph
  {et~al.}(2020{\natexlab{a}})\citenamefont {Sivarajah}, \citenamefont
  {Dilkes}, \citenamefont {Cowtan}, \citenamefont {Simmons}, \citenamefont
  {Edgington},\ and\ \citenamefont {Duncan}}]{tket2020}%
  \BibitemOpen
  \bibfield  {author} {\bibinfo {author} {\bibfnamefont {S.}~\bibnamefont
  {Sivarajah}}, \bibinfo {author} {\bibfnamefont {S.}~\bibnamefont {Dilkes}},
  \bibinfo {author} {\bibfnamefont {A.}~\bibnamefont {Cowtan}}, \bibinfo
  {author} {\bibfnamefont {W.}~\bibnamefont {Simmons}}, \bibinfo {author}
  {\bibfnamefont {A.}~\bibnamefont {Edgington}},\ and\ \bibinfo {author}
  {\bibfnamefont {R.}~\bibnamefont {Duncan}},\ }\href@noop {} {\bibfield
  {journal} {\bibinfo  {journal} {Quantum Science and Technology}\ }\textbf
  {\bibinfo {volume} {6}},\ \bibinfo {pages} {014003} (\bibinfo {year}
  {2020}{\natexlab{a}})}\BibitemShut {NoStop}%
\bibitem [{\citenamefont {Kliuchnikov}\ and\ \citenamefont
  {Maslov}(2013)}]{kliuchnikov2013optimization}%
  \BibitemOpen
  \bibfield  {author} {\bibinfo {author} {\bibfnamefont {V.}~\bibnamefont
  {Kliuchnikov}}\ and\ \bibinfo {author} {\bibfnamefont {D.}~\bibnamefont
  {Maslov}},\ }\href@noop {} {\bibfield  {journal} {\bibinfo  {journal}
  {Physical Review A}\ }\textbf {\bibinfo {volume} {88}},\ \bibinfo {pages}
  {052307} (\bibinfo {year} {2013})}\BibitemShut {NoStop}%
\bibitem [{\citenamefont {Abdessaied}\ \emph {et~al.}(2014)\citenamefont
  {Abdessaied}, \citenamefont {Soeken},\ and\ \citenamefont
  {Drechsler}}]{abdessaied2014quantum}%
  \BibitemOpen
  \bibfield  {author} {\bibinfo {author} {\bibfnamefont {N.}~\bibnamefont
  {Abdessaied}}, \bibinfo {author} {\bibfnamefont {M.}~\bibnamefont {Soeken}},\
  and\ \bibinfo {author} {\bibfnamefont {R.}~\bibnamefont {Drechsler}},\ }in\
  \href@noop {} {\emph {\bibinfo {booktitle} {Reversible Computation: 6th
  International Conference, RC 2014}}}\ (\bibinfo {organization} {Springer},\
  \bibinfo {year} {2014})\ pp.\ \bibinfo {pages} {149--162}\BibitemShut
  {NoStop}%
\bibitem [{\citenamefont {Nam}\ \emph {et~al.}(2018{\natexlab{a}})\citenamefont
  {Nam}, \citenamefont {Ross}, \citenamefont {Su}, \citenamefont {Childs},\
  and\ \citenamefont {Maslov}}]{nam2018automated}%
  \BibitemOpen
  \bibfield  {author} {\bibinfo {author} {\bibfnamefont {Y.}~\bibnamefont
  {Nam}}, \bibinfo {author} {\bibfnamefont {N.~J.}\ \bibnamefont {Ross}},
  \bibinfo {author} {\bibfnamefont {Y.}~\bibnamefont {Su}}, \bibinfo {author}
  {\bibfnamefont {A.~M.}\ \bibnamefont {Childs}},\ and\ \bibinfo {author}
  {\bibfnamefont {D.}~\bibnamefont {Maslov}},\ }\href@noop {} {\bibfield
  {journal} {\bibinfo  {journal} {npj Quantum Information}\ }\textbf {\bibinfo
  {volume} {4}},\ \bibinfo {pages} {23} (\bibinfo {year}
  {2018}{\natexlab{a}})}\BibitemShut {NoStop}%
\bibitem [{\citenamefont {Pointing}\ \emph {et~al.}(2021)\citenamefont
  {Pointing}, \citenamefont {Padon}, \citenamefont {Jia}, \citenamefont {Ma},
  \citenamefont {Hirth}, \citenamefont {Palsberg},\ and\ \citenamefont
  {Aiken}}]{pointing2021optimizing}%
  \BibitemOpen
  \bibfield  {author} {\bibinfo {author} {\bibfnamefont {J.}~\bibnamefont
  {Pointing}}, \bibinfo {author} {\bibfnamefont {O.}~\bibnamefont {Padon}},
  \bibinfo {author} {\bibfnamefont {Z.}~\bibnamefont {Jia}}, \bibinfo {author}
  {\bibfnamefont {H.}~\bibnamefont {Ma}}, \bibinfo {author} {\bibfnamefont
  {A.}~\bibnamefont {Hirth}}, \bibinfo {author} {\bibfnamefont
  {J.}~\bibnamefont {Palsberg}},\ and\ \bibinfo {author} {\bibfnamefont
  {A.}~\bibnamefont {Aiken}},\ }\href@noop {} {\bibinfo {title} {Quanto:
  Optimizing quantum circuits with automatic generation of circuit identities}}
  (\bibinfo {year} {2021}),\ \bibinfo {note} {arXiv:2111.11387}\BibitemShut
  {NoStop}%
\bibitem [{\citenamefont {Xu}\ \emph {et~al.}(2022)\citenamefont {Xu},
  \citenamefont {Li}, \citenamefont {Padon}, \citenamefont {Lin}, \citenamefont
  {Pointing}, \citenamefont {Hirth}, \citenamefont {Ma}, \citenamefont
  {Palsberg}, \citenamefont {Aiken}, \citenamefont {Acar} \emph
  {et~al.}}]{xu2022quartz}%
  \BibitemOpen
  \bibfield  {author} {\bibinfo {author} {\bibfnamefont {M.}~\bibnamefont
  {Xu}}, \bibinfo {author} {\bibfnamefont {Z.}~\bibnamefont {Li}}, \bibinfo
  {author} {\bibfnamefont {O.}~\bibnamefont {Padon}}, \bibinfo {author}
  {\bibfnamefont {S.}~\bibnamefont {Lin}}, \bibinfo {author} {\bibfnamefont
  {J.}~\bibnamefont {Pointing}}, \bibinfo {author} {\bibfnamefont
  {A.}~\bibnamefont {Hirth}}, \bibinfo {author} {\bibfnamefont
  {H.}~\bibnamefont {Ma}}, \bibinfo {author} {\bibfnamefont {J.}~\bibnamefont
  {Palsberg}}, \bibinfo {author} {\bibfnamefont {A.}~\bibnamefont {Aiken}},
  \bibinfo {author} {\bibfnamefont {U.~A.}\ \bibnamefont {Acar}}, \emph
  {et~al.},\ }in\ \href@noop {} {\emph {\bibinfo {booktitle} {Proceedings of
  the 43rd ACM SIGPLAN International Conference on Programming Language Design
  and Implementation}}}\ (\bibinfo {year} {2022})\ pp.\ \bibinfo {pages}
  {625--640}\BibitemShut {NoStop}%
\bibitem [{\citenamefont {Paykin}\ \emph
  {et~al.}(2023{\natexlab{a}})\citenamefont {Paykin}, \citenamefont {Schmitz},
  \citenamefont {Ibrahim}, \citenamefont {Wu},\ and\ \citenamefont
  {Matsuura}}]{PaykinSchmitz2023PCOAST}%
  \BibitemOpen
  \bibfield  {author} {\bibinfo {author} {\bibfnamefont {J.}~\bibnamefont
  {Paykin}}, \bibinfo {author} {\bibfnamefont {A.~T.}\ \bibnamefont {Schmitz}},
  \bibinfo {author} {\bibfnamefont {M.}~\bibnamefont {Ibrahim}}, \bibinfo
  {author} {\bibfnamefont {X.-C.}\ \bibnamefont {Wu}},\ and\ \bibinfo {author}
  {\bibfnamefont {A.~Y.}\ \bibnamefont {Matsuura}},\ }\href@noop {} {\bibinfo
  {title} {{PCOAST}: A {P}auli-based quantum circuit optimization framework
  ({E}xtended version)}} (\bibinfo {year} {2023}{\natexlab{a}}),\ \bibinfo
  {note} {arXiv}\BibitemShut {NoStop}%
\bibitem [{\citenamefont {Zhang}\ and\ \citenamefont {Chen}(2019)}]{Zhang2019}%
  \BibitemOpen
  \bibfield  {author} {\bibinfo {author} {\bibfnamefont {F.}~\bibnamefont
  {Zhang}}\ and\ \bibinfo {author} {\bibfnamefont {J.}~\bibnamefont {Chen}},\
  }\href {https://doi.org/10.48550/ARXIV.1903.12456} {\bibinfo {title}
  {Optimizing {T} gates in {Clifford+T} circuit as $\pi/4$ rotations around
  {Paulis}}} (\bibinfo {year} {2019}),\ \bibinfo {note} {arXiv:1903.12456},\
  \Eprint {https://arxiv.org/abs/1903.12456} {arXiv:1903.12456 [quant-ph]}
  \BibitemShut {NoStop}%
\bibitem [{\citenamefont {Sivarajah}\ \emph
  {et~al.}(2020{\natexlab{b}})\citenamefont {Sivarajah}, \citenamefont
  {Dilkes}, \citenamefont {Cowtan}, \citenamefont {Simmons}, \citenamefont
  {Edgington},\ and\ \citenamefont {Duncan}}]{Sivarajah_2021_tket}%
  \BibitemOpen
  \bibfield  {author} {\bibinfo {author} {\bibfnamefont {S.}~\bibnamefont
  {Sivarajah}}, \bibinfo {author} {\bibfnamefont {S.}~\bibnamefont {Dilkes}},
  \bibinfo {author} {\bibfnamefont {A.}~\bibnamefont {Cowtan}}, \bibinfo
  {author} {\bibfnamefont {W.}~\bibnamefont {Simmons}}, \bibinfo {author}
  {\bibfnamefont {A.}~\bibnamefont {Edgington}},\ and\ \bibinfo {author}
  {\bibfnamefont {R.}~\bibnamefont {Duncan}},\ }\href
  {https://doi.org/10.1088/2058-9565/ab8e92} {\bibfield  {journal} {\bibinfo
  {journal} {Quantum Science and Technology}\ }\textbf {\bibinfo {volume}
  {6}},\ \bibinfo {pages} {014003} (\bibinfo {year}
  {2020}{\natexlab{b}})}\BibitemShut {NoStop}%
\bibitem [{\citenamefont {Cowtan}\ \emph
  {et~al.}(2019{\natexlab{a}})\citenamefont {Cowtan}, \citenamefont {Dilkes},
  \citenamefont {Duncan}, \citenamefont {Simmons},\ and\ \citenamefont
  {Sivarajah}}]{cowtan2019phase}%
  \BibitemOpen
  \bibfield  {author} {\bibinfo {author} {\bibfnamefont {A.}~\bibnamefont
  {Cowtan}}, \bibinfo {author} {\bibfnamefont {S.}~\bibnamefont {Dilkes}},
  \bibinfo {author} {\bibfnamefont {R.}~\bibnamefont {Duncan}}, \bibinfo
  {author} {\bibfnamefont {W.}~\bibnamefont {Simmons}},\ and\ \bibinfo {author}
  {\bibfnamefont {S.}~\bibnamefont {Sivarajah}},\ }in\ \href@noop {} {\emph
  {\bibinfo {booktitle} {16th International Conference on Quantum Physics and
  Logic 2019}}}\ (\bibinfo {organization} {Open Publishing Association},\
  \bibinfo {year} {2019})\ pp.\ \bibinfo {pages} {213--228}\BibitemShut
  {NoStop}%
\bibitem [{\citenamefont {Lu}\ \emph {et~al.}(2021)\citenamefont {Lu},
  \citenamefont {Shen},\ and\ \citenamefont {Deng}}]{Lu2021}%
  \BibitemOpen
  \bibfield  {author} {\bibinfo {author} {\bibfnamefont {Z.}~\bibnamefont
  {Lu}}, \bibinfo {author} {\bibfnamefont {P.-X.}\ \bibnamefont {Shen}},\ and\
  \bibinfo {author} {\bibfnamefont {D.-L.}\ \bibnamefont {Deng}},\ }\href
  {https://doi.org/10.1103/PhysRevApplied.16.044039} {\bibfield  {journal}
  {\bibinfo  {journal} {Phys. Rev. Appl.}\ }\textbf {\bibinfo {volume} {16}},\
  \bibinfo {pages} {044039} (\bibinfo {year} {2021})}\BibitemShut {NoStop}%
\bibitem [{\citenamefont {Wecker}\ \emph {et~al.}(2015)\citenamefont {Wecker},
  \citenamefont {Hastings}, \citenamefont {Wiebe}, \citenamefont {Clark},
  \citenamefont {Nayak},\ and\ \citenamefont {Troyer}}]{wecker15_correlec}%
  \BibitemOpen
  \bibfield  {author} {\bibinfo {author} {\bibfnamefont {D.}~\bibnamefont
  {Wecker}}, \bibinfo {author} {\bibfnamefont {M.~B.}\ \bibnamefont
  {Hastings}}, \bibinfo {author} {\bibfnamefont {N.}~\bibnamefont {Wiebe}},
  \bibinfo {author} {\bibfnamefont {B.~K.}\ \bibnamefont {Clark}}, \bibinfo
  {author} {\bibfnamefont {C.}~\bibnamefont {Nayak}},\ and\ \bibinfo {author}
  {\bibfnamefont {M.}~\bibnamefont {Troyer}},\ }\href
  {https://doi.org/10.1103/PhysRevA.92.062318} {\bibfield  {journal} {\bibinfo
  {journal} {Phys. Rev. A}\ }\textbf {\bibinfo {volume} {92}},\ \bibinfo
  {pages} {062318} (\bibinfo {year} {2015})}\BibitemShut {NoStop}%
\bibitem [{\citenamefont {Abrams}\ and\ \citenamefont
  {Lloyd}(1997)}]{abrams97_fermion}%
  \BibitemOpen
  \bibfield  {author} {\bibinfo {author} {\bibfnamefont {D.~S.}\ \bibnamefont
  {Abrams}}\ and\ \bibinfo {author} {\bibfnamefont {S.}~\bibnamefont {Lloyd}},\
  }\href {https://doi.org/10.1103/PhysRevLett.79.2586} {\bibfield  {journal}
  {\bibinfo  {journal} {Phys. Rev. Lett.}\ }\textbf {\bibinfo {volume} {79}},\
  \bibinfo {pages} {2586} (\bibinfo {year} {1997})}\BibitemShut {NoStop}%
\bibitem [{\citenamefont {Nachman}\ \emph {et~al.}(2021)\citenamefont
  {Nachman}, \citenamefont {Provasoli}, \citenamefont {de~Jong},\ and\
  \citenamefont {Bauer}}]{Nachman21_hep}%
  \BibitemOpen
  \bibfield  {author} {\bibinfo {author} {\bibfnamefont {B.}~\bibnamefont
  {Nachman}}, \bibinfo {author} {\bibfnamefont {D.}~\bibnamefont {Provasoli}},
  \bibinfo {author} {\bibfnamefont {W.~A.}\ \bibnamefont {de~Jong}},\ and\
  \bibinfo {author} {\bibfnamefont {C.~W.}\ \bibnamefont {Bauer}},\ }\href
  {https://doi.org/10.1103/PhysRevLett.126.062001} {\bibfield  {journal}
  {\bibinfo  {journal} {Phys. Rev. Lett.}\ }\textbf {\bibinfo {volume} {126}},\
  \bibinfo {pages} {062001} (\bibinfo {year} {2021})}\BibitemShut {NoStop}%
\bibitem [{\citenamefont {Bauer}\ \emph {et~al.}(2020)\citenamefont {Bauer},
  \citenamefont {Bravyi}, \citenamefont {Motta},\ and\ \citenamefont
  {Chan}}]{Bauer20_chemrev}%
  \BibitemOpen
  \bibfield  {author} {\bibinfo {author} {\bibfnamefont {B.}~\bibnamefont
  {Bauer}}, \bibinfo {author} {\bibfnamefont {S.}~\bibnamefont {Bravyi}},
  \bibinfo {author} {\bibfnamefont {M.}~\bibnamefont {Motta}},\ and\ \bibinfo
  {author} {\bibfnamefont {G.~K.-L.}\ \bibnamefont {Chan}},\ }\href
  {https://doi.org/10.1021/acs.chemrev.9b00829} {\bibfield  {journal} {\bibinfo
   {journal} {Chemical Reviews}\ }\textbf {\bibinfo {volume} {120}},\ \bibinfo
  {pages} {12685} (\bibinfo {year} {2020})}\BibitemShut {NoStop}%
\bibitem [{\citenamefont {Ma}\ \emph {et~al.}(2020)\citenamefont {Ma},
  \citenamefont {Govoni},\ and\ \citenamefont {Galli}}]{Ma2020_mater}%
  \BibitemOpen
  \bibfield  {author} {\bibinfo {author} {\bibfnamefont {H.}~\bibnamefont
  {Ma}}, \bibinfo {author} {\bibfnamefont {M.}~\bibnamefont {Govoni}},\ and\
  \bibinfo {author} {\bibfnamefont {G.}~\bibnamefont {Galli}},\ }\bibfield
  {journal} {\bibinfo  {journal} {npj Computational Materials}\ }\textbf
  {\bibinfo {volume} {6}},\ \href {https://doi.org/10.1038/s41524-020-00353-z}
  {10.1038/s41524-020-00353-z} (\bibinfo {year} {2020})\BibitemShut {NoStop}%
\bibitem [{\citenamefont {Cao}\ \emph {et~al.}(2019)\citenamefont {Cao},
  \citenamefont {Romero}, \citenamefont {Olson}, \citenamefont {Degroote},
  \citenamefont {Johnson}, \citenamefont {Kieferová}, \citenamefont
  {Kivlichan}, \citenamefont {Menke}, \citenamefont {Peropadre}, \citenamefont
  {Sawaya}, \citenamefont {Sim}, \citenamefont {Veis},\ and\ \citenamefont
  {Aspuru-Guzik}}]{cao19_chemrev}%
  \BibitemOpen
  \bibfield  {author} {\bibinfo {author} {\bibfnamefont {Y.}~\bibnamefont
  {Cao}}, \bibinfo {author} {\bibfnamefont {J.}~\bibnamefont {Romero}},
  \bibinfo {author} {\bibfnamefont {J.~P.}\ \bibnamefont {Olson}}, \bibinfo
  {author} {\bibfnamefont {M.}~\bibnamefont {Degroote}}, \bibinfo {author}
  {\bibfnamefont {P.~D.}\ \bibnamefont {Johnson}}, \bibinfo {author}
  {\bibfnamefont {M.}~\bibnamefont {Kieferová}}, \bibinfo {author}
  {\bibfnamefont {I.~D.}\ \bibnamefont {Kivlichan}}, \bibinfo {author}
  {\bibfnamefont {T.}~\bibnamefont {Menke}}, \bibinfo {author} {\bibfnamefont
  {B.}~\bibnamefont {Peropadre}}, \bibinfo {author} {\bibfnamefont {N.~P.~D.}\
  \bibnamefont {Sawaya}}, \bibinfo {author} {\bibfnamefont {S.}~\bibnamefont
  {Sim}}, \bibinfo {author} {\bibfnamefont {L.}~\bibnamefont {Veis}},\ and\
  \bibinfo {author} {\bibfnamefont {A.}~\bibnamefont {Aspuru-Guzik}},\
  }\href@noop {} {\bibfield  {journal} {\bibinfo  {journal} {Chemical Reviews}\
  }\textbf {\bibinfo {volume} {2019}},\ \bibinfo {pages} {DOI:
  10.1021/acs.chemrev.8b00803} (\bibinfo {year} {2019})},\ \bibinfo {note}
  {arXiv:1812.09976}\BibitemShut {NoStop}%
\bibitem [{\citenamefont {McArdle}\ \emph {et~al.}(2020)\citenamefont
  {McArdle}, \citenamefont {Endo}, \citenamefont {Aspuru-Guzik}, \citenamefont
  {Benjamin},\ and\ \citenamefont {Yuan}}]{mcardle20_rev}%
  \BibitemOpen
  \bibfield  {author} {\bibinfo {author} {\bibfnamefont {S.}~\bibnamefont
  {McArdle}}, \bibinfo {author} {\bibfnamefont {S.}~\bibnamefont {Endo}},
  \bibinfo {author} {\bibfnamefont {A.}~\bibnamefont {Aspuru-Guzik}}, \bibinfo
  {author} {\bibfnamefont {S.~C.}\ \bibnamefont {Benjamin}},\ and\ \bibinfo
  {author} {\bibfnamefont {X.}~\bibnamefont {Yuan}},\ }\href
  {https://doi.org/10.1103/RevModPhys.92.015003} {\bibfield  {journal}
  {\bibinfo  {journal} {Rev. Mod. Phys.}\ }\textbf {\bibinfo {volume} {92}},\
  \bibinfo {pages} {015003} (\bibinfo {year} {2020})}\BibitemShut {NoStop}%
\bibitem [{\citenamefont {Elfving}\ \emph {et~al.}(2020)\citenamefont
  {Elfving}, \citenamefont {Broer}, \citenamefont {Webber}, \citenamefont
  {Gavartin}, \citenamefont {Halls}, \citenamefont {Lorton},\ and\
  \citenamefont {Bochevarov}}]{elfving20_indust}%
  \BibitemOpen
  \bibfield  {author} {\bibinfo {author} {\bibfnamefont {V.~E.}\ \bibnamefont
  {Elfving}}, \bibinfo {author} {\bibfnamefont {B.~W.}\ \bibnamefont {Broer}},
  \bibinfo {author} {\bibfnamefont {M.}~\bibnamefont {Webber}}, \bibinfo
  {author} {\bibfnamefont {J.}~\bibnamefont {Gavartin}}, \bibinfo {author}
  {\bibfnamefont {M.~D.}\ \bibnamefont {Halls}}, \bibinfo {author}
  {\bibfnamefont {K.~P.}\ \bibnamefont {Lorton}},\ and\ \bibinfo {author}
  {\bibfnamefont {A.}~\bibnamefont {Bochevarov}},\ }\href@noop {} {\bibfield
  {journal} {\bibinfo  {journal} {{a}rXiv}\ } (\bibinfo {year} {2020})},\
  \bibinfo {note} {{a}rXiv:2009.12472},\ \Eprint
  {https://arxiv.org/abs/arXiv:2009.12472} {arXiv:2009.12472} \BibitemShut
  {NoStop}%
\bibitem [{\citenamefont {McArdle}\ \emph {et~al.}(2019)\citenamefont
  {McArdle}, \citenamefont {Mayorov}, \citenamefont {Shan}, \citenamefont
  {Benjamin},\ and\ \citenamefont {Yuan}}]{mcardle19_qvibr}%
  \BibitemOpen
  \bibfield  {author} {\bibinfo {author} {\bibfnamefont {S.}~\bibnamefont
  {McArdle}}, \bibinfo {author} {\bibfnamefont {A.}~\bibnamefont {Mayorov}},
  \bibinfo {author} {\bibfnamefont {X.}~\bibnamefont {Shan}}, \bibinfo {author}
  {\bibfnamefont {S.}~\bibnamefont {Benjamin}},\ and\ \bibinfo {author}
  {\bibfnamefont {X.}~\bibnamefont {Yuan}},\ }\href
  {https://doi.org/10.1039/c9sc01313j} {\bibfield  {journal} {\bibinfo
  {journal} {Chem. Sci.}\ }\textbf {\bibinfo {volume} {10}},\ \bibinfo {pages}
  {5725} (\bibinfo {year} {2019})}\BibitemShut {NoStop}%
\bibitem [{\citenamefont {Sawaya}\ \emph {et~al.}(2021)\citenamefont {Sawaya},
  \citenamefont {Paesani},\ and\ \citenamefont {Tabor}}]{sawaya20_vibrspec}%
  \BibitemOpen
  \bibfield  {author} {\bibinfo {author} {\bibfnamefont {N.~P.}\ \bibnamefont
  {Sawaya}}, \bibinfo {author} {\bibfnamefont {F.}~\bibnamefont {Paesani}},\
  and\ \bibinfo {author} {\bibfnamefont {D.~P.}\ \bibnamefont {Tabor}},\
  }\href@noop {} {\bibfield  {journal} {\bibinfo  {journal} {Physical Review
  A}\ }\textbf {\bibinfo {volume} {104}},\ \bibinfo {pages} {062419} (\bibinfo
  {year} {2021})}\BibitemShut {NoStop}%
\bibitem [{\citenamefont {Ollitrault}\ \emph {et~al.}(2020)\citenamefont
  {Ollitrault}, \citenamefont {Baiardi}, \citenamefont {Reiher},\ and\
  \citenamefont {Tavernelli}}]{ollitrault20_reiher_qvibr}%
  \BibitemOpen
  \bibfield  {author} {\bibinfo {author} {\bibfnamefont {P.~J.}\ \bibnamefont
  {Ollitrault}}, \bibinfo {author} {\bibfnamefont {A.}~\bibnamefont {Baiardi}},
  \bibinfo {author} {\bibfnamefont {M.}~\bibnamefont {Reiher}},\ and\ \bibinfo
  {author} {\bibfnamefont {I.}~\bibnamefont {Tavernelli}},\ }\href
  {https://doi.org/10.1039/d0sc01908a} {\bibfield  {journal} {\bibinfo
  {journal} {Chem. Sci.}\ }\textbf {\bibinfo {volume} {11}},\ \bibinfo {pages}
  {6842} (\bibinfo {year} {2020})}\BibitemShut {NoStop}%
\bibitem [{\citenamefont {Suzuki}(1976)}]{suzuki76}%
  \BibitemOpen
  \bibfield  {author} {\bibinfo {author} {\bibfnamefont {M.}~\bibnamefont
  {Suzuki}},\ }\href {https://doi.org/10.1007/BF01609348} {\bibfield  {journal}
  {\bibinfo  {journal} {Communications in Mathematical Physics}\ }\textbf
  {\bibinfo {volume} {51}},\ \bibinfo {pages} {183} (\bibinfo {year}
  {1976})}\BibitemShut {NoStop}%
\bibitem [{\citenamefont {Berry}\ \emph {et~al.}(2015)\citenamefont {Berry},
  \citenamefont {Childs}, \citenamefont {Cleve}, \citenamefont {Kothari},\ and\
  \citenamefont {Somma}}]{Berry2015}%
  \BibitemOpen
  \bibfield  {author} {\bibinfo {author} {\bibfnamefont {D.~W.}\ \bibnamefont
  {Berry}}, \bibinfo {author} {\bibfnamefont {A.~M.}\ \bibnamefont {Childs}},
  \bibinfo {author} {\bibfnamefont {R.}~\bibnamefont {Cleve}}, \bibinfo
  {author} {\bibfnamefont {R.}~\bibnamefont {Kothari}},\ and\ \bibinfo {author}
  {\bibfnamefont {R.~D.}\ \bibnamefont {Somma}},\ }\href
  {https://doi.org/10.1103/PhysRevLett.114.090502} {\bibfield  {journal}
  {\bibinfo  {journal} {Physical Review Letters}\ }\textbf {\bibinfo {volume}
  {114}},\ \bibinfo {pages} {090502} (\bibinfo {year} {2015})}\BibitemShut
  {NoStop}%
\bibitem [{\citenamefont {Low}\ and\ \citenamefont {Chuang}(2017)}]{qsp2017}%
  \BibitemOpen
  \bibfield  {author} {\bibinfo {author} {\bibfnamefont {G.~H.}\ \bibnamefont
  {Low}}\ and\ \bibinfo {author} {\bibfnamefont {I.~L.}\ \bibnamefont
  {Chuang}},\ }\href {https://doi.org/10.1103/PhysRevLett.118.010501}
  {\bibfield  {journal} {\bibinfo  {journal} {Phys. Rev. Lett.}\ }\textbf
  {\bibinfo {volume} {118}},\ \bibinfo {pages} {010501} (\bibinfo {year}
  {2017})}\BibitemShut {NoStop}%
\bibitem [{\citenamefont {Low}\ and\ \citenamefont
  {Chuang}(2019)}]{low2019_qubitization}%
  \BibitemOpen
  \bibfield  {author} {\bibinfo {author} {\bibfnamefont {G.~H.}\ \bibnamefont
  {Low}}\ and\ \bibinfo {author} {\bibfnamefont {I.~L.}\ \bibnamefont
  {Chuang}},\ }\href {https://doi.org/10.22331/q-2019-07-12-163} {\bibfield
  {journal} {\bibinfo  {journal} {Quantum}\ }\textbf {\bibinfo {volume} {3}},\
  \bibinfo {pages} {163} (\bibinfo {year} {2019})}\BibitemShut {NoStop}%
\bibitem [{\citenamefont {Childs}\ \emph {et~al.}(2019)\citenamefont {Childs},
  \citenamefont {Ostrander},\ and\ \citenamefont {Su}}]{Childs2019_random}%
  \BibitemOpen
  \bibfield  {author} {\bibinfo {author} {\bibfnamefont {A.~M.}\ \bibnamefont
  {Childs}}, \bibinfo {author} {\bibfnamefont {A.}~\bibnamefont {Ostrander}},\
  and\ \bibinfo {author} {\bibfnamefont {Y.}~\bibnamefont {Su}},\ }\href
  {https://doi.org/10.22331/q-2019-09-02-182} {\bibfield  {journal} {\bibinfo
  {journal} {Quantum}\ }\textbf {\bibinfo {volume} {3}},\ \bibinfo {pages}
  {182} (\bibinfo {year} {2019})}\BibitemShut {NoStop}%
\bibitem [{\citenamefont {Campbell}(2019)}]{Campbell2019_random}%
  \BibitemOpen
  \bibfield  {author} {\bibinfo {author} {\bibfnamefont {E.}~\bibnamefont
  {Campbell}},\ }\href {https://doi.org/10.1103/PhysRevLett.123.070503}
  {\bibfield  {journal} {\bibinfo  {journal} {Physical Review Letters}\
  }\textbf {\bibinfo {volume} {123}},\ \bibinfo {pages} {070503} (\bibinfo
  {year} {2019})}\BibitemShut {NoStop}%
\bibitem [{\citenamefont {Berry}\ and\ \citenamefont
  {Childs}(2012)}]{berry15_blackbox}%
  \BibitemOpen
  \bibfield  {author} {\bibinfo {author} {\bibfnamefont {D.~W.}\ \bibnamefont
  {Berry}}\ and\ \bibinfo {author} {\bibfnamefont {A.~M.}\ \bibnamefont
  {Childs}},\ }\href@noop {} {\bibfield  {journal} {\bibinfo  {journal}
  {Quantum Info. Comput.}\ }\textbf {\bibinfo {volume} {12}},\ \bibinfo {pages}
  {29–62} (\bibinfo {year} {2012})}\BibitemShut {NoStop}%
\bibitem [{\citenamefont {{Berry}}\ \emph {et~al.}(2015)\citenamefont
  {{Berry}}, \citenamefont {{Childs}},\ and\ \citenamefont
  {{Kothari}}}]{berry15_hamsim_qwalk}%
  \BibitemOpen
  \bibfield  {author} {\bibinfo {author} {\bibfnamefont {D.~W.}\ \bibnamefont
  {{Berry}}}, \bibinfo {author} {\bibfnamefont {A.~M.}\ \bibnamefont
  {{Childs}}},\ and\ \bibinfo {author} {\bibfnamefont {R.}~\bibnamefont
  {{Kothari}}},\ }in\ \href {https://doi.org/10.1109/FOCS.2015.54} {\emph
  {\bibinfo {booktitle} {2015 IEEE 56th Annual Symposium on Foundations of
  Computer Science}}}\ (\bibinfo {year} {2015})\ pp.\ \bibinfo {pages}
  {792--809}\BibitemShut {NoStop}%
\bibitem [{\citenamefont {Childs}\ \emph {et~al.}(2021)\citenamefont {Childs},
  \citenamefont {Su}, \citenamefont {Tran}, \citenamefont {Wiebe},\ and\
  \citenamefont {Zhu}}]{childs19_theoryoftrotter}%
  \BibitemOpen
  \bibfield  {author} {\bibinfo {author} {\bibfnamefont {A.~M.}\ \bibnamefont
  {Childs}}, \bibinfo {author} {\bibfnamefont {Y.}~\bibnamefont {Su}}, \bibinfo
  {author} {\bibfnamefont {M.~C.}\ \bibnamefont {Tran}}, \bibinfo {author}
  {\bibfnamefont {N.}~\bibnamefont {Wiebe}},\ and\ \bibinfo {author}
  {\bibfnamefont {S.}~\bibnamefont {Zhu}},\ }\href
  {https://doi.org/10.1103/PhysRevX.11.011020} {\bibfield  {journal} {\bibinfo
  {journal} {Phys. Rev. X}\ }\textbf {\bibinfo {volume} {11}},\ \bibinfo
  {pages} {011020} (\bibinfo {year} {2021})}\BibitemShut {NoStop}%
\bibitem [{Note1()}]{Note1}%
  \BibitemOpen
  \bibinfo {note} {As a practical matter, no implementation can truly realize
  all arbitrary single-qubit rotations natively. However, one only needs at
  minimum the Clifford $H$ gate and the non-Clifford $T$ gate, at which point
  any other rotation can be achieved with arbitrary precision \cite
  {Nielsenbook}. Thus it is merely a theoretical convenience to assume
  arbitrary rotations from the outset.}\BibitemShut {Stop}%
\bibitem [{\citenamefont {Barenco}\ \emph {et~al.}(1995)\citenamefont
  {Barenco}, \citenamefont {Bennett}, \citenamefont {Cleve}, \citenamefont
  {DiVincenzo}, \citenamefont {Margolus}, \citenamefont {Shor}, \citenamefont
  {Sleator}, \citenamefont {Smolin},\ and\ \citenamefont
  {Weinfurter}}]{Barenco1995}%
  \BibitemOpen
  \bibfield  {author} {\bibinfo {author} {\bibfnamefont {A.}~\bibnamefont
  {Barenco}}, \bibinfo {author} {\bibfnamefont {C.~H.}\ \bibnamefont
  {Bennett}}, \bibinfo {author} {\bibfnamefont {R.}~\bibnamefont {Cleve}},
  \bibinfo {author} {\bibfnamefont {D.~P.}\ \bibnamefont {DiVincenzo}},
  \bibinfo {author} {\bibfnamefont {N.}~\bibnamefont {Margolus}}, \bibinfo
  {author} {\bibfnamefont {P.}~\bibnamefont {Shor}}, \bibinfo {author}
  {\bibfnamefont {T.}~\bibnamefont {Sleator}}, \bibinfo {author} {\bibfnamefont
  {J.~A.}\ \bibnamefont {Smolin}},\ and\ \bibinfo {author} {\bibfnamefont
  {H.}~\bibnamefont {Weinfurter}},\ }\href
  {https://doi.org/10.1103/PhysRevA.52.3457} {\bibfield  {journal} {\bibinfo
  {journal} {Phys. Rev. A}\ }\textbf {\bibinfo {volume} {52}},\ \bibinfo
  {pages} {3457} (\bibinfo {year} {1995})}\BibitemShut {NoStop}%
\bibitem [{\citenamefont {Nielsen}\ and\ \citenamefont
  {Chuang}(2011)}]{Nielsenbook}%
  \BibitemOpen
  \bibfield  {author} {\bibinfo {author} {\bibfnamefont {M.~A.}\ \bibnamefont
  {Nielsen}}\ and\ \bibinfo {author} {\bibfnamefont {I.~L.}\ \bibnamefont
  {Chuang}},\ }\href@noop {} {\emph {\bibinfo {title} {Quantum Computation and
  Quantum Information: 10th Anniversary Edition}}},\ \bibinfo {edition} {10th}\
  ed.\ (\bibinfo  {publisher} {Cambridge University Press},\ \bibinfo {address}
  {New York, NY, USA},\ \bibinfo {year} {2011})\BibitemShut {NoStop}%
\bibitem [{\citenamefont {Fagan}\ and\ \citenamefont
  {Duncan}(2018)}]{Fagan2018}%
  \BibitemOpen
  \bibfield  {author} {\bibinfo {author} {\bibfnamefont {A.}~\bibnamefont
  {Fagan}}\ and\ \bibinfo {author} {\bibfnamefont {R.}~\bibnamefont {Duncan}},\
  }in\ \href@noop {} {\emph {\bibinfo {booktitle} {QPL}}}\ (\bibinfo {year}
  {2018})\BibitemShut {NoStop}%
\bibitem [{\citenamefont {Nam}\ \emph {et~al.}(2018{\natexlab{b}})\citenamefont
  {Nam}, \citenamefont {Ross}, \citenamefont {Su}, \citenamefont {Childs},\
  and\ \citenamefont {Maslov}}]{Nam2018}%
  \BibitemOpen
  \bibfield  {author} {\bibinfo {author} {\bibfnamefont {Y.}~\bibnamefont
  {Nam}}, \bibinfo {author} {\bibfnamefont {N.~J.}\ \bibnamefont {Ross}},
  \bibinfo {author} {\bibfnamefont {Y.}~\bibnamefont {Su}}, \bibinfo {author}
  {\bibfnamefont {A.~M.}\ \bibnamefont {Childs}},\ and\ \bibinfo {author}
  {\bibfnamefont {D.}~\bibnamefont {Maslov}},\ }\href
  {https://doi.org/10.1038/s41534-018-0072-4} {\bibfield  {journal} {\bibinfo
  {journal} {npj Quantum Information}\ }\textbf {\bibinfo {volume} {4}},\
  \bibinfo {pages} {23} (\bibinfo {year} {2018}{\natexlab{b}})}\BibitemShut
  {NoStop}%
\bibitem [{\citenamefont {Nash}\ \emph {et~al.}(2020)\citenamefont {Nash},
  \citenamefont {Gheorghiu},\ and\ \citenamefont {Mosca}}]{Nash2020}%
  \BibitemOpen
  \bibfield  {author} {\bibinfo {author} {\bibfnamefont {B.}~\bibnamefont
  {Nash}}, \bibinfo {author} {\bibfnamefont {V.}~\bibnamefont {Gheorghiu}},\
  and\ \bibinfo {author} {\bibfnamefont {M.}~\bibnamefont {Mosca}},\ }\href
  {https://doi.org/10.1088/2058-9565/ab79b1} {\bibfield  {journal} {\bibinfo
  {journal} {Quantum Science and Technology}\ }\textbf {\bibinfo {volume}
  {5}},\ \bibinfo {pages} {025010} (\bibinfo {year} {2020})}\BibitemShut
  {NoStop}%
\bibitem [{\citenamefont {{Maslov}}\ and\ \citenamefont
  {{Roetteler}}(2018)}]{Maslov2018}%
  \BibitemOpen
  \bibfield  {author} {\bibinfo {author} {\bibfnamefont {D.}~\bibnamefont
  {{Maslov}}}\ and\ \bibinfo {author} {\bibfnamefont {M.}~\bibnamefont
  {{Roetteler}}},\ }\href {https://doi.org/10.1109/TIT.2018.2825602} {\bibfield
   {journal} {\bibinfo  {journal} {IEEE Transactions on Information Theory}\
  }\textbf {\bibinfo {volume} {64}},\ \bibinfo {pages} {4729} (\bibinfo {year}
  {2018})}\BibitemShut {NoStop}%
\bibitem [{\citenamefont {Aaronson}\ and\ \citenamefont
  {Gottesman}(2004)}]{Aaronson2004}%
  \BibitemOpen
  \bibfield  {author} {\bibinfo {author} {\bibfnamefont {S.}~\bibnamefont
  {Aaronson}}\ and\ \bibinfo {author} {\bibfnamefont {D.}~\bibnamefont
  {Gottesman}},\ }\href {https://doi.org/10.1103/PhysRevA.70.052328} {\bibfield
   {journal} {\bibinfo  {journal} {Phys. Rev. A}\ }\textbf {\bibinfo {volume}
  {70}},\ \bibinfo {pages} {052328} (\bibinfo {year} {2004})}\BibitemShut
  {NoStop}%
\bibitem [{\citenamefont {Cowtan}\ \emph
  {et~al.}(2019{\natexlab{b}})\citenamefont {Cowtan}, \citenamefont {Dilkes},
  \citenamefont {Duncan}, \citenamefont {Simmons},\ and\ \citenamefont
  {Sivarajah}}]{Cowtan2019}%
  \BibitemOpen
  \bibfield  {author} {\bibinfo {author} {\bibfnamefont {A.}~\bibnamefont
  {Cowtan}}, \bibinfo {author} {\bibfnamefont {S.}~\bibnamefont {Dilkes}},
  \bibinfo {author} {\bibfnamefont {R.}~\bibnamefont {Duncan}}, \bibinfo
  {author} {\bibfnamefont {W.}~\bibnamefont {Simmons}},\ and\ \bibinfo {author}
  {\bibfnamefont {S.}~\bibnamefont {Sivarajah}},\ }\href
  {https://doi.org/10.4204/EPTCS.318.13} {\bibfield  {journal} {\bibinfo
  {journal} {Proceedings 16th International Conference on Quantum Physics and
  Logic, {QPL} 2019, Chapman University, Orange, CA, USA, June 10-14, 2019}\
  }\bibinfo {series} {{EPTCS}},\ \textbf {\bibinfo {volume} {318}},\ \bibinfo
  {pages} {213} (\bibinfo {year} {2019}{\natexlab{b}})}\BibitemShut {NoStop}%
\bibitem [{\citenamefont {Motta}\ \emph {et~al.}(2021)\citenamefont {Motta},
  \citenamefont {Ye}, \citenamefont {McClean}, \citenamefont {Li},
  \citenamefont {Minnich}, \citenamefont {Babbush},\ and\ \citenamefont
  {Chan}}]{Motta2018}%
  \BibitemOpen
  \bibfield  {author} {\bibinfo {author} {\bibfnamefont {M.}~\bibnamefont
  {Motta}}, \bibinfo {author} {\bibfnamefont {E.}~\bibnamefont {Ye}}, \bibinfo
  {author} {\bibfnamefont {J.~R.}\ \bibnamefont {McClean}}, \bibinfo {author}
  {\bibfnamefont {Z.}~\bibnamefont {Li}}, \bibinfo {author} {\bibfnamefont
  {A.~J.}\ \bibnamefont {Minnich}}, \bibinfo {author} {\bibfnamefont
  {R.}~\bibnamefont {Babbush}},\ and\ \bibinfo {author} {\bibfnamefont
  {G.~K.-L.}\ \bibnamefont {Chan}},\ }\href@noop {} {\bibfield  {journal}
  {\bibinfo  {journal} {npj Quantum Information}\ }\textbf {\bibinfo {volume}
  {7}},\ \bibinfo {pages} {1} (\bibinfo {year} {2021})}\BibitemShut {NoStop}%
\bibitem [{\citenamefont {Hastings}\ \emph {et~al.}(2015)\citenamefont
  {Hastings}, \citenamefont {Wecker}, \citenamefont {Bauer},\ and\
  \citenamefont {Troyer}}]{Hastings2015}%
  \BibitemOpen
  \bibfield  {author} {\bibinfo {author} {\bibfnamefont {M.~B.}\ \bibnamefont
  {Hastings}}, \bibinfo {author} {\bibfnamefont {D.}~\bibnamefont {Wecker}},
  \bibinfo {author} {\bibfnamefont {B.}~\bibnamefont {Bauer}},\ and\ \bibinfo
  {author} {\bibfnamefont {M.}~\bibnamefont {Troyer}},\ }\href@noop {}
  {\bibfield  {journal} {\bibinfo  {journal} {Quantum Info. Comput.}\ }\textbf
  {\bibinfo {volume} {15}},\ \bibinfo {pages} {121} (\bibinfo {year}
  {2015})}\BibitemShut {NoStop}%
\bibitem [{\citenamefont {Li}\ \emph {et~al.}(2022)\citenamefont {Li},
  \citenamefont {Wu}, \citenamefont {Shi}, \citenamefont {Javadi-Abhari},
  \citenamefont {Ding},\ and\ \citenamefont {Xie}}]{2022paulihedral}%
  \BibitemOpen
  \bibfield  {author} {\bibinfo {author} {\bibfnamefont {G.}~\bibnamefont
  {Li}}, \bibinfo {author} {\bibfnamefont {A.}~\bibnamefont {Wu}}, \bibinfo
  {author} {\bibfnamefont {Y.}~\bibnamefont {Shi}}, \bibinfo {author}
  {\bibfnamefont {A.}~\bibnamefont {Javadi-Abhari}}, \bibinfo {author}
  {\bibfnamefont {Y.}~\bibnamefont {Ding}},\ and\ \bibinfo {author}
  {\bibfnamefont {Y.}~\bibnamefont {Xie}},\ }in\ \href
  {https://doi.org/10.1145/3503222.3507715} {\emph {\bibinfo {booktitle}
  {Proceedings of the 27th ACM International Conference on Architectural
  Support for Programming Languages and Operating Systems}}},\ \bibinfo {series
  and number} {ASPLOS '22}\ (\bibinfo  {publisher} {Association for Computing
  Machinery},\ \bibinfo {address} {New York, NY, USA},\ \bibinfo {year}
  {2022})\ p.\ \bibinfo {pages} {554–569}\BibitemShut {NoStop}%
\bibitem [{\citenamefont {Bodmann}\ \emph {et~al.}(2009)\citenamefont
  {Bodmann}, \citenamefont {Le}, \citenamefont {Reza}, \citenamefont {Tobin},\
  and\ \citenamefont {Tomforde}}]{Bernhard2009}%
  \BibitemOpen
  \bibfield  {author} {\bibinfo {author} {\bibfnamefont {B.~G.}\ \bibnamefont
  {Bodmann}}, \bibinfo {author} {\bibfnamefont {M.}~\bibnamefont {Le}},
  \bibinfo {author} {\bibfnamefont {L.}~\bibnamefont {Reza}}, \bibinfo {author}
  {\bibfnamefont {M.}~\bibnamefont {Tobin}},\ and\ \bibinfo {author}
  {\bibfnamefont {M.}~\bibnamefont {Tomforde}},\ }\href@noop {} {\bibinfo
  {title} {Frame theory for binary vector spaces}} (\bibinfo {year} {2009}),\
  \bibinfo {note} {arXiv: 0906.3467}\BibitemShut {NoStop}%
\bibitem [{\citenamefont {Terhal}(2015)}]{Terhal2015}%
  \BibitemOpen
  \bibfield  {author} {\bibinfo {author} {\bibfnamefont {B.~M.}\ \bibnamefont
  {Terhal}},\ }\href {https://doi.org/10.1103/RevModPhys.87.307} {\bibfield
  {journal} {\bibinfo  {journal} {Rev. Mod. Phys.}\ }\textbf {\bibinfo {volume}
  {87}},\ \bibinfo {pages} {307} (\bibinfo {year} {2015})}\BibitemShut
  {NoStop}%
\bibitem [{\citenamefont {{Gottesman}}(1997)}]{Gottesman1997}%
  \BibitemOpen
  \bibfield  {author} {\bibinfo {author} {\bibfnamefont {D.}~\bibnamefont
  {{Gottesman}}},\ }\emph {\bibinfo {title} {Stabilizer codes and quantum error
  correction}},\ \href@noop {} {Ph.D. thesis},\ \bibinfo  {school} {California
  Institute of Technology} (\bibinfo {year} {1997})\BibitemShut {NoStop}%
\bibitem [{\citenamefont {Schmitz}(2019)}]{Schmitz2019}%
  \BibitemOpen
  \bibfield  {author} {\bibinfo {author} {\bibfnamefont {A.~T.}\ \bibnamefont
  {Schmitz}},\ }\href
  {https://doi.org/https://doi.org/10.1016/j.aop.2019.167927} {\bibfield
  {journal} {\bibinfo  {journal} {Annals of Physics}\ }\textbf {\bibinfo
  {volume} {410}},\ \bibinfo {pages} {167927} (\bibinfo {year}
  {2019})}\BibitemShut {NoStop}%
\bibitem [{\citenamefont {Tolar}(2018)}]{Tolar2018}%
  \BibitemOpen
  \bibfield  {author} {\bibinfo {author} {\bibfnamefont {J.}~\bibnamefont
  {Tolar}},\ }\href {https://doi.org/10.1088/1742-6596/1071/1/012022}
  {\bibfield  {journal} {\bibinfo  {journal} {Journal of Physics: Conference
  Series}\ }\textbf {\bibinfo {volume} {1071}},\ \bibinfo {pages} {012022}
  (\bibinfo {year} {2018})}\BibitemShut {NoStop}%
\bibitem [{Note2()}]{Note2}%
  \BibitemOpen
  \bibinfo {note} {Recall that unitarity is defined in the context of an inner
  product space $(\protect \mathcal V, \mathinner {\langle {,}\rangle })$ where
  a unitary operator is defined as any linear operator $U$ such that for all
  $u,v \in \protect \mathcal V$, $\mathinner {\langle {U(u), U(v)}\rangle } =
  \mathinner {\langle {u,v}\rangle }$. This then implies the usual notion of
  unitarity that $U^\dagger = U^{-1}$, where $\dagger $ signifies the adjoint
  with respects to $\mathinner {\langle {,}\rangle }$.}\BibitemShut {Stop}%
\bibitem [{Note3()}]{Note3}%
  \BibitemOpen
  \bibinfo {note} {A reasonable question one might have is why we don't include
  an extra bit in the Pauli space to represent the sign. Though this is
  possible in principle, this causes two problems. By definition, this would
  make the binary two-form $\lambda $ degenerate in this extended space, and
  thus could not be used to form any frame, let alone a Pauli frame. The second
  reason is that two-qubit Clifford gates such as CX are no longer linear in
  this extended space. Moreover, this non-linearity is not a practical problem
  for tracking the sign. See Supplemental Material for details.}\BibitemShut
  {Stop}%
\bibitem [{Note4()}]{Note4}%
  \BibitemOpen
  \bibinfo {note} {The choice of CX over CZ is not consequential as we'll see
  in Section \ref {sec:trottercycle}.}\BibitemShut {Stop}%
\bibitem [{Note5()}]{Note5}%
  \BibitemOpen
  \bibinfo {note} {There is some intuition for why this is. Consider the
  analogy of rotating 3D vectors. Rotation as an operation on some 3D vector
  can be view as either the direct or ``forward'' rotation of the vector while
  fixing coordinates, or the inverse or ``backwards'' rotation of the
  coordinates while fixing the vector. As we shall see in a moment, the methods
  we describe below are analogous to the former. We think of the Pauli
  operators as being fixed, while Clifford operations transform the
  ``coordinates'' of the space around it.}\BibitemShut {Stop}%
\bibitem [{\citenamefont {Paykin}\ \emph
  {et~al.}(2023{\natexlab{b}})\citenamefont {Paykin}, \citenamefont {Schmitz},
  \citenamefont {Ibrahim}, \citenamefont {Wu},\ and\ \citenamefont
  {Matsuura}}]{2023Paykin}%
  \BibitemOpen
  \bibfield  {author} {\bibinfo {author} {\bibfnamefont {J.}~\bibnamefont
  {Paykin}}, \bibinfo {author} {\bibfnamefont {A.~T.}\ \bibnamefont {Schmitz}},
  \bibinfo {author} {\bibfnamefont {M.}~\bibnamefont {Ibrahim}}, \bibinfo
  {author} {\bibfnamefont {X.-C.}\ \bibnamefont {Wu}},\ and\ \bibinfo {author}
  {\bibfnamefont {A.~Y.}\ \bibnamefont {Matsuura}},\ }\href {arXiv} {\bibinfo
  {title} {{PCOAST}: A {P}auli-based quantum circuit optimization framework
  (extended version)}} (\bibinfo {year} {2023}{\natexlab{b}})\BibitemShut
  {NoStop}%
\bibitem [{Note6()}]{Note6}%
  \BibitemOpen
  \bibinfo {note} {It should be noted that $\protect \text {\protect
  \normalfont Supp}$ is not a distance function in the usual sense for one
  major reason: the two entries of its domain are not in the same space. Still,
  Proposition \ref {prop1} provides all the analogous properties to an actual
  distance function.}\BibitemShut {Stop}%
\bibitem [{Note7()}]{Note7}%
  \BibitemOpen
  \bibinfo {note} {Note that retracing does not undo the work done by the
  former step because we are not reversing the signs of the rotation
  angles.}\BibitemShut {Stop}%
\bibitem [{\citenamefont {McClean}\ \emph {et~al.}(2020)\citenamefont
  {McClean}, \citenamefont {Rubin}, \citenamefont {Sung}, \citenamefont
  {Kivlichan}, \citenamefont {Bonet-Monroig}, \citenamefont {Cao},
  \citenamefont {Dai}, \citenamefont {Fried}, \citenamefont {Gidney},
  \citenamefont {Gimby}, \citenamefont {Gokhale}, \citenamefont {Häner},
  \citenamefont {Hardikar}, \citenamefont {Havl{\'{\i}}{\v{c}}ek},
  \citenamefont {Higgott}, \citenamefont {Huang}, \citenamefont {Izaac},
  \citenamefont {Jiang}, \citenamefont {Liu}, \citenamefont {McArdle},
  \citenamefont {Neeley}, \citenamefont {O'Brien}, \citenamefont {O'Gorman},
  \citenamefont {Ozfidan}, \citenamefont {Radin}, \citenamefont {Romero},
  \citenamefont {Sawaya}, \citenamefont {Senjean}, \citenamefont {Setia},
  \citenamefont {Sim}, \citenamefont {Steiger}, \citenamefont {Steudtner},
  \citenamefont {Sun}, \citenamefont {Sun}, \citenamefont {Wang}, \citenamefont
  {Zhang},\ and\ \citenamefont {Babbush}}]{McClean_2020}%
  \BibitemOpen
  \bibfield  {author} {\bibinfo {author} {\bibfnamefont {J.~R.}\ \bibnamefont
  {McClean}}, \bibinfo {author} {\bibfnamefont {N.~C.}\ \bibnamefont {Rubin}},
  \bibinfo {author} {\bibfnamefont {K.~J.}\ \bibnamefont {Sung}}, \bibinfo
  {author} {\bibfnamefont {I.~D.}\ \bibnamefont {Kivlichan}}, \bibinfo {author}
  {\bibfnamefont {X.}~\bibnamefont {Bonet-Monroig}}, \bibinfo {author}
  {\bibfnamefont {Y.}~\bibnamefont {Cao}}, \bibinfo {author} {\bibfnamefont
  {C.}~\bibnamefont {Dai}}, \bibinfo {author} {\bibfnamefont {E.~S.}\
  \bibnamefont {Fried}}, \bibinfo {author} {\bibfnamefont {C.}~\bibnamefont
  {Gidney}}, \bibinfo {author} {\bibfnamefont {B.}~\bibnamefont {Gimby}},
  \bibinfo {author} {\bibfnamefont {P.}~\bibnamefont {Gokhale}}, \bibinfo
  {author} {\bibfnamefont {T.}~\bibnamefont {Häner}}, \bibinfo {author}
  {\bibfnamefont {T.}~\bibnamefont {Hardikar}}, \bibinfo {author}
  {\bibfnamefont {V.}~\bibnamefont {Havl{\'{\i}}{\v{c}}ek}}, \bibinfo {author}
  {\bibfnamefont {O.}~\bibnamefont {Higgott}}, \bibinfo {author} {\bibfnamefont
  {C.}~\bibnamefont {Huang}}, \bibinfo {author} {\bibfnamefont
  {J.}~\bibnamefont {Izaac}}, \bibinfo {author} {\bibfnamefont
  {Z.}~\bibnamefont {Jiang}}, \bibinfo {author} {\bibfnamefont
  {X.}~\bibnamefont {Liu}}, \bibinfo {author} {\bibfnamefont {S.}~\bibnamefont
  {McArdle}}, \bibinfo {author} {\bibfnamefont {M.}~\bibnamefont {Neeley}},
  \bibinfo {author} {\bibfnamefont {T.}~\bibnamefont {O'Brien}}, \bibinfo
  {author} {\bibfnamefont {B.}~\bibnamefont {O'Gorman}}, \bibinfo {author}
  {\bibfnamefont {I.}~\bibnamefont {Ozfidan}}, \bibinfo {author} {\bibfnamefont
  {M.~D.}\ \bibnamefont {Radin}}, \bibinfo {author} {\bibfnamefont
  {J.}~\bibnamefont {Romero}}, \bibinfo {author} {\bibfnamefont {N.~P.~D.}\
  \bibnamefont {Sawaya}}, \bibinfo {author} {\bibfnamefont {B.}~\bibnamefont
  {Senjean}}, \bibinfo {author} {\bibfnamefont {K.}~\bibnamefont {Setia}},
  \bibinfo {author} {\bibfnamefont {S.}~\bibnamefont {Sim}}, \bibinfo {author}
  {\bibfnamefont {D.~S.}\ \bibnamefont {Steiger}}, \bibinfo {author}
  {\bibfnamefont {M.}~\bibnamefont {Steudtner}}, \bibinfo {author}
  {\bibfnamefont {Q.}~\bibnamefont {Sun}}, \bibinfo {author} {\bibfnamefont
  {W.}~\bibnamefont {Sun}}, \bibinfo {author} {\bibfnamefont {D.}~\bibnamefont
  {Wang}}, \bibinfo {author} {\bibfnamefont {F.}~\bibnamefont {Zhang}},\ and\
  \bibinfo {author} {\bibfnamefont {R.}~\bibnamefont {Babbush}},\ }\href
  {https://doi.org/10.1088/2058-9565/ab8ebc} {\bibfield  {journal} {\bibinfo
  {journal} {Quantum Science and Technology}\ }\textbf {\bibinfo {volume}
  {5}},\ \bibinfo {pages} {034014} (\bibinfo {year} {2020})}\BibitemShut
  {NoStop}%
\bibitem [{\citenamefont {Sawaya}\ \emph {et~al.}(2020)\citenamefont {Sawaya},
  \citenamefont {Menke}, \citenamefont {Kyaw}, \citenamefont {Johri},
  \citenamefont {Aspuru-Guzik},\ and\ \citenamefont
  {Guerreschi}}]{sawaya19_dlev}%
  \BibitemOpen
  \bibfield  {author} {\bibinfo {author} {\bibfnamefont {N.~P.~D.}\
  \bibnamefont {Sawaya}}, \bibinfo {author} {\bibfnamefont {T.}~\bibnamefont
  {Menke}}, \bibinfo {author} {\bibfnamefont {T.~H.}\ \bibnamefont {Kyaw}},
  \bibinfo {author} {\bibfnamefont {S.}~\bibnamefont {Johri}}, \bibinfo
  {author} {\bibfnamefont {A.}~\bibnamefont {Aspuru-Guzik}},\ and\ \bibinfo
  {author} {\bibfnamefont {G.~G.}\ \bibnamefont {Guerreschi}},\ }\bibfield
  {journal} {\bibinfo  {journal} {npj Quantum Information}\ }\textbf {\bibinfo
  {volume} {6}},\ \href {https://doi.org/10.1038/s41534-020-0278-0}
  {10.1038/s41534-020-0278-0} (\bibinfo {year} {2020})\BibitemShut {NoStop}%
\bibitem [{\citenamefont {Bravyi}\ and\ \citenamefont {Kitaev}(2002)}]{bk02}%
  \BibitemOpen
  \bibfield  {author} {\bibinfo {author} {\bibfnamefont {S.~B.}\ \bibnamefont
  {Bravyi}}\ and\ \bibinfo {author} {\bibfnamefont {A.~Y.}\ \bibnamefont
  {Kitaev}},\ }\href {https://doi.org/10.1006/aphy.2002.6254} {\bibfield
  {journal} {\bibinfo  {journal} {Annals of Physics}\ }\textbf {\bibinfo
  {volume} {298}},\ \bibinfo {pages} {210} (\bibinfo {year}
  {2002})}\BibitemShut {NoStop}%
\bibitem [{\citenamefont {Tranter}\ \emph {et~al.}(2015)\citenamefont
  {Tranter}, \citenamefont {Sofia}, \citenamefont {Seeley}, \citenamefont
  {Kaicher}, \citenamefont {McClean}, \citenamefont {Babbush}, \citenamefont
  {Coveney}, \citenamefont {Mintert}, \citenamefont {Wilhelm},\ and\
  \citenamefont {Love}}]{tranter15_bk}%
  \BibitemOpen
  \bibfield  {author} {\bibinfo {author} {\bibfnamefont {A.}~\bibnamefont
  {Tranter}}, \bibinfo {author} {\bibfnamefont {S.}~\bibnamefont {Sofia}},
  \bibinfo {author} {\bibfnamefont {J.}~\bibnamefont {Seeley}}, \bibinfo
  {author} {\bibfnamefont {M.}~\bibnamefont {Kaicher}}, \bibinfo {author}
  {\bibfnamefont {J.}~\bibnamefont {McClean}}, \bibinfo {author} {\bibfnamefont
  {R.}~\bibnamefont {Babbush}}, \bibinfo {author} {\bibfnamefont {P.~V.}\
  \bibnamefont {Coveney}}, \bibinfo {author} {\bibfnamefont {F.}~\bibnamefont
  {Mintert}}, \bibinfo {author} {\bibfnamefont {F.}~\bibnamefont {Wilhelm}},\
  and\ \bibinfo {author} {\bibfnamefont {P.~J.}\ \bibnamefont {Love}},\ }\href
  {https://doi.org/10.1002/qua.24969} {\bibfield  {journal} {\bibinfo
  {journal} {International Journal of Quantum Chemistry}\ }\textbf {\bibinfo
  {volume} {115}},\ \bibinfo {pages} {1431} (\bibinfo {year}
  {2015})}\BibitemShut {NoStop}%
\bibitem [{\citenamefont {Helgaker}\ \emph {et~al.}(2000)\citenamefont
  {Helgaker}, \citenamefont {J{\o}rgensen},\ and\ \citenamefont
  {Olsen}}]{helgaaker_book}%
  \BibitemOpen
  \bibfield  {author} {\bibinfo {author} {\bibfnamefont {T.}~\bibnamefont
  {Helgaker}}, \bibinfo {author} {\bibfnamefont {P.}~\bibnamefont
  {J{\o}rgensen}},\ and\ \bibinfo {author} {\bibfnamefont {J.}~\bibnamefont
  {Olsen}},\ }\href@noop {} {\emph {\bibinfo {title} {Molecular
  Electronic-Structure Theory}}}\ (\bibinfo  {publisher} {John Wiley \& Sons},\
  \bibinfo {year} {2000})\BibitemShut {NoStop}%
\bibitem [{\citenamefont {Whitfield}\ \emph {et~al.}(2011)\citenamefont
  {Whitfield}, \citenamefont {Biamonte},\ and\ \citenamefont
  {Aspuru-Guzik}}]{Whitfield2011}%
  \BibitemOpen
  \bibfield  {author} {\bibinfo {author} {\bibfnamefont {J.~D.}\ \bibnamefont
  {Whitfield}}, \bibinfo {author} {\bibfnamefont {J.}~\bibnamefont
  {Biamonte}},\ and\ \bibinfo {author} {\bibfnamefont {A.}~\bibnamefont
  {Aspuru-Guzik}},\ }\href {https://doi.org/10.1080/00268976.2011.552441}
  {\bibfield  {journal} {\bibinfo  {journal} {Molecular Physics}\ }\textbf
  {\bibinfo {volume} {109}},\ \bibinfo {pages} {735} (\bibinfo {year}
  {2011})}\BibitemShut {NoStop}%
\bibitem [{\citenamefont {McClean}\ \emph {et~al.}(2014)\citenamefont
  {McClean}, \citenamefont {Babbush}, \citenamefont {Love},\ and\ \citenamefont
  {Aspuru-Guzik}}]{McClean2014_jpcl}%
  \BibitemOpen
  \bibfield  {author} {\bibinfo {author} {\bibfnamefont {J.~R.}\ \bibnamefont
  {McClean}}, \bibinfo {author} {\bibfnamefont {R.}~\bibnamefont {Babbush}},
  \bibinfo {author} {\bibfnamefont {P.~J.}\ \bibnamefont {Love}},\ and\
  \bibinfo {author} {\bibfnamefont {A.}~\bibnamefont {Aspuru-Guzik}},\ }\href
  {https://doi.org/10.1021/jz501649m} {\bibfield  {journal} {\bibinfo
  {journal} {The Journal of Physical Chemistry Letters}\ }\textbf {\bibinfo
  {volume} {5}},\ \bibinfo {pages} {4368} (\bibinfo {year} {2014})}\BibitemShut
  {NoStop}%
\bibitem [{\citenamefont {Babbush}\ \emph {et~al.}(2015)\citenamefont
  {Babbush}, \citenamefont {McClean}, \citenamefont {Wecker}, \citenamefont
  {Aspuru-Guzik},\ and\ \citenamefont {Wiebe}}]{babbush15_pra}%
  \BibitemOpen
  \bibfield  {author} {\bibinfo {author} {\bibfnamefont {R.}~\bibnamefont
  {Babbush}}, \bibinfo {author} {\bibfnamefont {J.}~\bibnamefont {McClean}},
  \bibinfo {author} {\bibfnamefont {D.}~\bibnamefont {Wecker}}, \bibinfo
  {author} {\bibfnamefont {A.}~\bibnamefont {Aspuru-Guzik}},\ and\ \bibinfo
  {author} {\bibfnamefont {N.}~\bibnamefont {Wiebe}},\ }\href
  {https://doi.org/10.1103/PhysRevA.91.022311} {\bibfield  {journal} {\bibinfo
  {journal} {Phys. Rev. A}\ }\textbf {\bibinfo {volume} {91}},\ \bibinfo
  {pages} {022311} (\bibinfo {year} {2015})}\BibitemShut {NoStop}%
\bibitem [{\citenamefont {Hanwell}\ \emph {et~al.}(2012)\citenamefont
  {Hanwell}, \citenamefont {Curtis}, \citenamefont {Lonie}, \citenamefont
  {Vandermeersch}, \citenamefont {Zurek},\ and\ \citenamefont
  {Hutchison}}]{avogadro2012}%
  \BibitemOpen
  \bibfield  {author} {\bibinfo {author} {\bibfnamefont {M.~D.}\ \bibnamefont
  {Hanwell}}, \bibinfo {author} {\bibfnamefont {D.~E.}\ \bibnamefont {Curtis}},
  \bibinfo {author} {\bibfnamefont {D.~C.}\ \bibnamefont {Lonie}}, \bibinfo
  {author} {\bibfnamefont {T.}~\bibnamefont {Vandermeersch}}, \bibinfo {author}
  {\bibfnamefont {E.}~\bibnamefont {Zurek}},\ and\ \bibinfo {author}
  {\bibfnamefont {G.~R.}\ \bibnamefont {Hutchison}},\ }\bibfield  {journal}
  {\bibinfo  {journal} {Journal of Cheminformatics}\ }\textbf {\bibinfo
  {volume} {4}},\ \href {https://doi.org/10.1186/1758-2946-4-17}
  {10.1186/1758-2946-4-17} (\bibinfo {year} {2012})\BibitemShut {NoStop}%
\bibitem [{\citenamefont {Rappe}\ \emph {et~al.}(1992)\citenamefont {Rappe},
  \citenamefont {Casewit}, \citenamefont {Colwell}, \citenamefont {Goddard},\
  and\ \citenamefont {Skiff}}]{rappe92_uff}%
  \BibitemOpen
  \bibfield  {author} {\bibinfo {author} {\bibfnamefont {A.~K.}\ \bibnamefont
  {Rappe}}, \bibinfo {author} {\bibfnamefont {C.~J.}\ \bibnamefont {Casewit}},
  \bibinfo {author} {\bibfnamefont {K.~S.}\ \bibnamefont {Colwell}}, \bibinfo
  {author} {\bibfnamefont {W.~A.}\ \bibnamefont {Goddard}},\ and\ \bibinfo
  {author} {\bibfnamefont {W.~M.}\ \bibnamefont {Skiff}},\ }\href
  {https://doi.org/10.1021/ja00051a040} {\bibfield  {journal} {\bibinfo
  {journal} {Journal of the American Chemical Society}\ }\textbf {\bibinfo
  {volume} {114}},\ \bibinfo {pages} {10024} (\bibinfo {year} {1992})},\
  \Eprint {https://arxiv.org/abs/https://doi.org/10.1021/ja00051a040}
  {https://doi.org/10.1021/ja00051a040} \BibitemShut {NoStop}%
\bibitem [{\citenamefont {Parrish}\ \emph {et~al.}(2017)\citenamefont
  {Parrish}, \citenamefont {Burns}, \citenamefont {Smith}, \citenamefont
  {Simmonett}, \citenamefont {DePrince}, \citenamefont {Hohenstein},
  \citenamefont {Bozkaya}, \citenamefont {Sokolov}, \citenamefont {Di~Remigio},
  \citenamefont {Richard}, \citenamefont {Gonthier}, \citenamefont {James},
  \citenamefont {McAlexander}, \citenamefont {Kumar}, \citenamefont {Saitow},
  \citenamefont {Wang}, \citenamefont {Pritchard}, \citenamefont {Verma},
  \citenamefont {Schaefer}, \citenamefont {Patkowski}, \citenamefont {King},
  \citenamefont {Valeev}, \citenamefont {Evangelista}, \citenamefont {Turney},
  \citenamefont {Crawford},\ and\ \citenamefont {Sherrill}}]{psi4_2017}%
  \BibitemOpen
  \bibfield  {author} {\bibinfo {author} {\bibfnamefont {R.~M.}\ \bibnamefont
  {Parrish}}, \bibinfo {author} {\bibfnamefont {L.~A.}\ \bibnamefont {Burns}},
  \bibinfo {author} {\bibfnamefont {D.~G.~A.}\ \bibnamefont {Smith}}, \bibinfo
  {author} {\bibfnamefont {A.~C.}\ \bibnamefont {Simmonett}}, \bibinfo {author}
  {\bibfnamefont {A.~E.}\ \bibnamefont {DePrince}}, \bibinfo {author}
  {\bibfnamefont {E.~G.}\ \bibnamefont {Hohenstein}}, \bibinfo {author}
  {\bibfnamefont {U.}~\bibnamefont {Bozkaya}}, \bibinfo {author} {\bibfnamefont
  {A.~Y.}\ \bibnamefont {Sokolov}}, \bibinfo {author} {\bibfnamefont
  {R.}~\bibnamefont {Di~Remigio}}, \bibinfo {author} {\bibfnamefont {R.~M.}\
  \bibnamefont {Richard}}, \bibinfo {author} {\bibfnamefont {J.~F.}\
  \bibnamefont {Gonthier}}, \bibinfo {author} {\bibfnamefont {A.~M.}\
  \bibnamefont {James}}, \bibinfo {author} {\bibfnamefont {H.~R.}\ \bibnamefont
  {McAlexander}}, \bibinfo {author} {\bibfnamefont {A.}~\bibnamefont {Kumar}},
  \bibinfo {author} {\bibfnamefont {M.}~\bibnamefont {Saitow}}, \bibinfo
  {author} {\bibfnamefont {X.}~\bibnamefont {Wang}}, \bibinfo {author}
  {\bibfnamefont {B.~P.}\ \bibnamefont {Pritchard}}, \bibinfo {author}
  {\bibfnamefont {P.}~\bibnamefont {Verma}}, \bibinfo {author} {\bibfnamefont
  {H.~F.}\ \bibnamefont {Schaefer}}, \bibinfo {author} {\bibfnamefont
  {K.}~\bibnamefont {Patkowski}}, \bibinfo {author} {\bibfnamefont {R.~A.}\
  \bibnamefont {King}}, \bibinfo {author} {\bibfnamefont {E.~F.}\ \bibnamefont
  {Valeev}}, \bibinfo {author} {\bibfnamefont {F.~A.}\ \bibnamefont
  {Evangelista}}, \bibinfo {author} {\bibfnamefont {J.~M.}\ \bibnamefont
  {Turney}}, \bibinfo {author} {\bibfnamefont {T.~D.}\ \bibnamefont
  {Crawford}},\ and\ \bibinfo {author} {\bibfnamefont {C.~D.}\ \bibnamefont
  {Sherrill}},\ }\href {https://doi.org/10.1021/acs.jctc.7b00174} {\bibfield
  {journal} {\bibinfo  {journal} {Journal of Chemical Theory and Computation}\
  }\textbf {\bibinfo {volume} {13}},\ \bibinfo {pages} {3185} (\bibinfo {year}
  {2017})},\ \bibinfo {note} {pMID: 28489372},\ \Eprint
  {https://arxiv.org/abs/https://doi.org/10.1021/acs.jctc.7b00174}
  {https://doi.org/10.1021/acs.jctc.7b00174} \BibitemShut {NoStop}%
\bibitem [{\citenamefont {Fisher}\ \emph {et~al.}(1989)\citenamefont {Fisher},
  \citenamefont {Weichman}, \citenamefont {Grinstein},\ and\ \citenamefont
  {Fisher}}]{fisher89}%
  \BibitemOpen
  \bibfield  {author} {\bibinfo {author} {\bibfnamefont {M.~P.~A.}\
  \bibnamefont {Fisher}}, \bibinfo {author} {\bibfnamefont {P.~B.}\
  \bibnamefont {Weichman}}, \bibinfo {author} {\bibfnamefont {G.}~\bibnamefont
  {Grinstein}},\ and\ \bibinfo {author} {\bibfnamefont {D.~S.}\ \bibnamefont
  {Fisher}},\ }\href {https://doi.org/10.1103/PhysRevB.40.546} {\bibfield
  {journal} {\bibinfo  {journal} {Phys. Rev. B}\ }\textbf {\bibinfo {volume}
  {40}},\ \bibinfo {pages} {546} (\bibinfo {year} {1989})}\BibitemShut
  {NoStop}%
\bibitem [{\citenamefont {Bloch}\ \emph {et~al.}(2008)\citenamefont {Bloch},
  \citenamefont {Dalibard},\ and\ \citenamefont {Zwerger}}]{bloch_review}%
  \BibitemOpen
  \bibfield  {author} {\bibinfo {author} {\bibfnamefont {I.}~\bibnamefont
  {Bloch}}, \bibinfo {author} {\bibfnamefont {J.}~\bibnamefont {Dalibard}},\
  and\ \bibinfo {author} {\bibfnamefont {W.}~\bibnamefont {Zwerger}},\ }\href
  {https://doi.org/10.1103/RevModPhys.80.885} {\bibfield  {journal} {\bibinfo
  {journal} {Rev. Mod. Phys.}\ }\textbf {\bibinfo {volume} {80}},\ \bibinfo
  {pages} {885} (\bibinfo {year} {2008})}\BibitemShut {NoStop}%
\bibitem [{\citenamefont {Sawaya}\ and\ \citenamefont
  {Huh}(2019)}]{sawaya19_vibronic}%
  \BibitemOpen
  \bibfield  {author} {\bibinfo {author} {\bibfnamefont {N.~P.~D.}\
  \bibnamefont {Sawaya}}\ and\ \bibinfo {author} {\bibfnamefont
  {J.}~\bibnamefont {Huh}},\ }\href
  {https://doi.org/10.1021/acs.jpclett.9b01117} {\bibfield  {journal} {\bibinfo
   {journal} {The Journal of Physical Chemistry Letters}\ }\textbf {\bibinfo
  {volume} {10}},\ \bibinfo {pages} {3586} (\bibinfo {year}
  {2019})}\BibitemShut {NoStop}%
\bibitem [{Note8()}]{Note8}%
  \BibitemOpen
  \bibinfo {note} {This does not include the return path as we assume
  retracing.}\BibitemShut {Stop}%
\bibitem [{\citenamefont {Schmitz}\ \emph
  {et~al.}(2023{\natexlab{a}})\citenamefont {Schmitz}, \citenamefont {Ibrahim},
  \citenamefont {Sawaya}, \citenamefont {Guerreschi}, \citenamefont {Paykin},
  \citenamefont {Wu},\ and\ \citenamefont {Matsuura}}]{schmitz2023PCOAST}%
  \BibitemOpen
  \bibfield  {author} {\bibinfo {author} {\bibfnamefont {A.~T.}\ \bibnamefont
  {Schmitz}}, \bibinfo {author} {\bibfnamefont {M.}~\bibnamefont {Ibrahim}},
  \bibinfo {author} {\bibfnamefont {N.~P.~D.}\ \bibnamefont {Sawaya}}, \bibinfo
  {author} {\bibfnamefont {G.~G.}\ \bibnamefont {Guerreschi}}, \bibinfo
  {author} {\bibfnamefont {J.}~\bibnamefont {Paykin}}, \bibinfo {author}
  {\bibfnamefont {X.-C.}\ \bibnamefont {Wu}},\ and\ \bibinfo {author}
  {\bibfnamefont {A.~Y.}\ \bibnamefont {Matsuura}},\ }\href@noop {} {\bibinfo
  {title} {Optimization at the interface of unitary and non-unitary quantum
  operations in {PCOAST}}} (\bibinfo {year} {2023}{\natexlab{a}}),\ \bibinfo
  {note} {arXiv:2305.09843},\ \Eprint {https://arxiv.org/abs/2305.09843}
  {arXiv:2305.09843 [quant-ph]} \BibitemShut {NoStop}%
\bibitem [{\citenamefont {Khalate}\ \emph {et~al.}(2022)\citenamefont
  {Khalate}, \citenamefont {Wu}, \citenamefont {Premaratne}, \citenamefont
  {Hogaboam}, \citenamefont {Holmes}, \citenamefont {Schmitz}, \citenamefont
  {Guerreschi}, \citenamefont {Zou},\ and\ \citenamefont
  {Matsuura}}]{Khalate2022}%
  \BibitemOpen
  \bibfield  {author} {\bibinfo {author} {\bibfnamefont {P.}~\bibnamefont
  {Khalate}}, \bibinfo {author} {\bibfnamefont {X.-C.}\ \bibnamefont {Wu}},
  \bibinfo {author} {\bibfnamefont {S.}~\bibnamefont {Premaratne}}, \bibinfo
  {author} {\bibfnamefont {J.}~\bibnamefont {Hogaboam}}, \bibinfo {author}
  {\bibfnamefont {A.}~\bibnamefont {Holmes}}, \bibinfo {author} {\bibfnamefont
  {A.}~\bibnamefont {Schmitz}}, \bibinfo {author} {\bibfnamefont {G.~G.}\
  \bibnamefont {Guerreschi}}, \bibinfo {author} {\bibfnamefont
  {X.}~\bibnamefont {Zou}},\ and\ \bibinfo {author} {\bibfnamefont {A.~Y.}\
  \bibnamefont {Matsuura}},\ }\href {https://doi.org/10.48550/ARXIV.2202.11142}
  {\bibinfo {title} {An {LLVM}-based {C++} compiler toolchain for variational
  hybrid quantum-classical algorithms and quantum accelerators}} (\bibinfo
  {year} {2022}),\ \bibinfo {note} {arXiv:2202.11142}\BibitemShut {NoStop}%
\bibitem [{\citenamefont {Schmitz}\ \emph
  {et~al.}(2023{\natexlab{b}})\citenamefont {Schmitz}, \citenamefont {Paykin},\
  and\ \citenamefont {Matsuura}}]{schmitz2023functional}%
  \BibitemOpen
  \bibfield  {author} {\bibinfo {author} {\bibfnamefont {A.}~\bibnamefont
  {Schmitz}}, \bibinfo {author} {\bibfnamefont {J.}~\bibnamefont {Paykin}},\
  and\ \bibinfo {author} {\bibfnamefont {A.}~\bibnamefont {Matsuura}},\
  }\href@noop {} {\bibfield  {journal} {\bibinfo  {journal} {Bulletin of the
  American Physical Society}\ } (\bibinfo {year}
  {2023}{\natexlab{b}})}\BibitemShut {NoStop}%
\bibitem [{Note9()}]{Note9}%
  \BibitemOpen
  \bibinfo {note} {Note this method relies on the fact that no process will try
  to schedule a rotation on the same qubit as any another. This is guaranteed
  by two facts: 1) no Pauli terms are duplicated, and thus each process has a
  distinct list of Pauli operator terms. 2) each pass through the main loop
  only applies one TQE gate. It should be clear that the application of a TQE
  gate only provides two new possible rotations not available in the previous
  frame. Furthermore, these are alway on distinct qubits. Together, one can see
  that aside for the initial case (as was handled at the beginning by the rank
  0 process), each pass through the main loop can not result in two rotation
  gates being scheduled on the same qubit. \label {fn1}}\BibitemShut {NoStop}%
\end{thebibliography}%

\end{document}